\documentclass[%
reprint,
superscriptaddress,
preprintnumbers,
 amsmath,amssymb,
 aps,
]{revtex4-1}

\usepackage{graphicx}
\usepackage{dcolumn}
\usepackage{bm}
\usepackage{hyperref}

\usepackage{float}
\usepackage[utf8]{inputenc} 
\usepackage{xcolor}
\usepackage{axodraw2}

\begin{document}

\preprint{CAVENDISH-HEP-16-19}

\title{Phenomenology of a Higgs triplet model at future $e^{+}e^{-}$ colliders\\}

\author{Sylvain Blunier}
\email{sblunier@uc.cl}
\affiliation{Instituto de F\'isica, Pontificia Universidad Cat\'olica de Chile, Av. Vicu\~na Mackenna 4860, Santiago, Chile}

\author{Giovanna Cottin}
\email{gfc24@cam.ac.uk}
\affiliation{Cavendish Laboratory, University of Cambridge, J.J. Thomson Ave, Cambridge CB3 0HE, UK}

\author{Marco A. D\'iaz}
\email{mad@susy.fis.puc.cl}
\affiliation{Instituto de F\'isica, Pontificia Universidad Cat\'olica de Chile, Av. Vicu\~na Mackenna 4860, Santiago, Chile}%

\author{Benjamin Koch}
\email{bkoch@fis.puc.cl}
\affiliation{Instituto de F\'isica, Pontificia Universidad Cat\'olica de Chile, Av. Vicu\~na Mackenna 4860, Santiago, Chile}%

\date{\today}

\begin{abstract}
In this work, we investigate the prospects of future $e^{+}e^{-}$ colliders in 
testing a Higgs triplet model with a scalar triplet and a scalar singlet under $SU(2)$.
The parameters of the model are fixed so that the lightest $CP-$even 
state corresponds to the Higgs particle observed at the LHC at around $125$ GeV.
This study investigates if the second heaviest $CP-$even, the heaviest $CP-$odd 
and the singly charged states can be observed at existing and future colliders by 
computing their accessible production and decay channels. 
In general, the LHC is not well equipped to produce a Higgs boson which is 
not mainly doublet-like, so we turn our focus to lepton colliders.
We find distinctive features of this model in cases when the second 
heaviest $CP-$even Higgs is triplet-like, singlet-like or a mixture. These 
features could distinguish the model from other scenarios at future $e^{+}e^{-}$ colliders.
\end{abstract}

\maketitle

\section{Introduction} 
The discovery of the Higgs boson at the LHC~\cite{Aad:2012tfa,Chatrchyan:2012xdj} confirms the particle 
content of the Standard Model (SM) of particle physics. Still one of the main beyond the SM puzzles remains 
neutrino mass generation. Several extensions to the SM Higgs sector that give a mass term to neutrinos
involve the spontaneous violation of lepton number via the vacuum expectation value of 
an $SU(2)$ singlet (for a review, see Ref.~\cite{Valle:1990pka}). A common feature of these models is 
the presence of a massless goldstone boson, the Majoron $J$. 

We investigate the phenomenology of a Higgs triplet model (HTM) of the kind mentioned above that has a scalar 
singlet and a scalar triplet under $SU(2)$, in addition to a $SU(2)$ scalar doublet. The model was originally 
proposed in~\cite{Schechter:1981cv}, where the authors defined it as the ``123'' HTM. Once the 
triplet field acquires a vacuum expectation value (vev), a neutrino mass term is 
generated. The parameters in the neutrino sector include the vev of the triplet and the 
Yukawa couplings between the two-component fermion $SU(2)$ doublet, including charged leptons and 
majorana neutrinos, and the triplet field. In this work, we study the collider phenomenology 
of the ``123" model, which is almost decoupled from its neutrino sector~\footnote{The connexion
between the neutrino sector of the model and collider physics arises via the decays 
of the doubly charged Higgs (arising from the triplet) to charged leptons,
as these decays involve the same Yukawas above mentioned.}. This is why we don't discuss experimental
constrains on neutrino masses and mixing angles, which are beyond the scope of this paper and which we leave 
for a future work. Models in which neutrino masses arise from the 
interaction with a triplet field have also been discussed extensively in 
the literature~\cite{Schechter:1980gr,Accomando:2006ga,Chun:2003ej,Nishiura:2009yd,*Akeroyd:2007zv,*Garayoa:2007fw,*Dev:2013ff,Cheng:1980qt}. 

The phenomenology of ``123'' models was studied before in~\cite{Diaz:1998zg,Akeroyd:2010eg}, paying 
particular attention to the consistency of the presence of the Majoron with experimental data. The Majoron is 
mainly singlet in this model, so its interaction with gauge bosons such as the $Z$ is negligible, making its existence 
fully consistent with collider data. This is in contrast to what happens in models with spontaneous violation of
lepton number without the singlet field~\cite{Gelmini:1980re}, which are excluded. 

A characteristic signature of models with Higgs triplets is the existence of a 
doubly charged scalar ($\Delta^{\pm\pm}$), in addition to the existence of a 
tree-level $H^{\pm}W^{\mp}Z$ vertex, where $H^{\pm}$ is a singly charged 
Higgs~\cite{Accomando:2006ga}. The LHC collider phenomenology of a doubly charged scalar in Higgs triplet 
models (in particular the ``23'' HTM, without the singlet field) has been discussed 
in~\cite{Chun:2003ej,Chiang:2012dk,*Akeroyd:2011ir,*Akeroyd:2011zza,*Akeroyd:2010ip,*Akeroyd:2009hb,*Perez:2008ha,*Akeroyd:2005gt}. Production of doubly charged scalars at $e^{+}e^{-}$ colliders has
also been studied in the literature as probes of Higgs triplet 
models~\cite{Shen:2015ora,*Cao:2014xba,*Shen:2014rpa,*Cao:2014roa,*Yagyu:2014aaa}, the 
Georgi–Machacek model~\cite{Yu:2016skv,*Chiang:2015rva,*Cheung:1994rp,*Godbole:1994np} and left-right 
symmetric models~\cite{Barenboim:1996pt}, which have a similar phenomenology. 

The phenomenology of the neutral scalar sector in Higgs triplet models has been less studied than the charged sector. 
Production and decays of the neutral Higgs bosons in the ``23'' HTM, was studied  
in~\cite{Arbabifar:2012bd,Akeroyd:2012nd,*Akeroyd:2010je}. Associated production of the charged and neutral 
Higgs at the ILC was studied in~\cite{Zhang:2016lad,Shen:2015pih}. In particular for the ``123'' HTM of 
interest in this paper, only discovery prospects at colliders were 
discussed in~\cite{Diaz:1998zg} and a fermiophobic Higgs was studied in~\cite{Akeroyd:2010eg}. 

The collider phenomenology of neutral and singly charged Higgs bosons 
in the HTM has received much less attention in the literature than 
the doubly charged Higgs. In addition, the phenomenology of the doubly charged Higgs 
depends directly on neutrino physics we are not evaluating 
at this time (as noticed earlier), so we focus on the neutral sector 
and singly charged Higgs of the ``123'' HTM. 

In this paper, we study the production and decay of the next to heaviest 
neutral $CP-$even Higgs $h_{2}$, the $CP-$odd Higgs $A$ and the singly charged 
Higgs $H^{\pm}$ of the ``123" HTM. We extend the work in Refs.~\cite{Diaz:1998zg,Akeroyd:2010eg} by identifying the 
lightest state in the $CP-$even neutral sector, $h_{1}$, as the SM-like Higgs discovered at 
the LHC. This rules out the fermiophobic SM-like Higgs boson scenario 
described in~\cite{Diaz:1998zg}. Constrains are imposed on the parameter space of the model in order to retain 
the SM-like Higgs properties. In particular, we define $h_{1}$ to be mainly doublet and fix its 
mass to be $m_{h_{1}}\approx 125$ GeV. We also identify 
the necessary constrains on the parameters of the scalar potential to suppress its decays to 
Majorons, so that its invisible decay width is negligible. 

We identify three characteristic benchmarks of the model related to the composition of $h_{2}$. $h_{2}$
can be mainly singlet, mainly triplet or a mixture. Note that $h_2$ can not be mainly a doublet 
since this is reserved for the SM like Higgs-boson. We compute production 
cross-sections and decays in these three benchmarks. We find that the 
main 2-body production mode for $h_{2}$ is associated production with a $CP-$odd state $A$ and note 
that cross-sections are in general larger when $A$ is produced on-shell. Production of
$A$ may be observable at CLIC when produced in association with an $h_{2}$ or 
$h_{3}$ (the heaviest $CP-$even Higgs), depending on the benchmark. The singly charged Higgs 
boson $H^{+}$ is potentially observable at CLIC when produced in association with another $H^{-}$.
Decay rates of $h_{2}$ to fermions are suppressed. Invisible decays of $h_{2}$ to 
Majorons can be very important, depending on the benchmark. Decays of $A\rightarrow h_{i} Z$, with $i=1,2$ or
$A\rightarrow t\bar{t}$ dominate, depending on the benchmark. The decays of $H^{\pm}\rightarrow h_{1}W^{\pm}$ 
dominate in all three benchmarks.

The paper is organized as follows. In Section~\ref{model} we introduce the 
model under study. Section~\ref{ParamRestrictions} describes our restrictions and scan over the parameter 
space. In Section~\ref{LHCprod} we comment on the low production cross-section of 
the new heavy Higgs of this model at the LHC. Section~\ref{eplusProd} describes production of $h_{2}$, $A$ and $H^{\pm}$
at future $e^{+}e^{-}$ colliders, while in Section~\ref{decays} we comment on the decay 
phenomenology of the model. We briefly comment on the most promising channels for discovery 
in Section~\ref{Channels}. After a summary and conclusions in Section~\ref{close} we define the relevant Feynman rules 
in Appendix~\ref{FeynmanRules}, for easy reference by the reader.

\section{The Model}
\label{model}

The model under consideration was introduced in Ref.~\cite{Schechter:1981cv} and studied further in 
Refs.~\cite{Diaz:1998zg,Akeroyd:2010eg}. The scalar sector includes a singlet $\sigma$ with lepton number 
$L_\sigma=2$ and hypercharge $Y_\sigma=0$, a doublet $\phi$ with lepton number $L_\phi=0$ and
hypercharge $Y_\phi=-1$, and a triplet $\Delta$ with lepton number $L_\Delta=-2$ and hypercharge
$Y_\Delta=2$. The notation we use is,
\begin{eqnarray}
\sigma &=& \frac{1}{\sqrt{2}}(v_\sigma+\chi_\sigma+i\varphi_\sigma),
\nonumber\\
\phi &=& \left( \begin{array}{c} \frac{1}{\sqrt{2}}(v_\phi+\chi_\phi+i\varphi_\phi) \\ \phi^- \end{array} 
\right),
\nonumber\\
\Delta &=& \left( \begin{array}{cc} \frac{1}{\sqrt{2}}(v_\Delta+\chi_\Delta+i\varphi_\Delta) & 
\Delta^+/\sqrt{2} \\ \Delta^+/\sqrt{2} & \Delta^{++} \end{array} \right),
\end{eqnarray}
where $v_\sigma$, $v_\phi$, $v_\Delta$ are the vacuum expectation values (vev) of the neutral components of 
each scalar field. 
The presence of the triplet allows to have a term that can give 
mass to neutrinos~\cite{Schechter:1980gr,Accomando:2006ga,Cheng:1980qt}.

Following the notation of \cite{Diaz:1998zg}, the scalar potential can be written as

\begin{align}
V(\sigma,\phi,\Delta) &= \mu_1^2 \sigma^\dagger \sigma + \mu_2^2 \phi^\dagger \phi
+ \mu_3^2 {\mathrm{Tr}} ( \Delta^\dagger \Delta ) + \lambda_1 (\phi^\dagger \phi)^2\nonumber \\
&+\lambda_2 \big[{\mathrm{Tr}} ( \Delta^\dagger \Delta )\big]^2 +\lambda_3 (\phi^\dagger \phi) {\mathrm{Tr}} ( \Delta^\dagger \Delta )
\nonumber\\ 
&+ \lambda_4 {\mathrm{Tr}} ( \Delta^\dagger \Delta \Delta^\dagger \Delta )
+ \lambda_5 ( \phi^\dagger \Delta^\dagger \Delta \phi ) + \beta_1 (\sigma^\dagger \sigma)^2\nonumber \\
&+ \beta_2 (\phi^\dagger \phi) (\sigma^\dagger \sigma)
+ \beta_3 {\mathrm{Tr}} ( \Delta^\dagger \Delta ) (\sigma^\dagger \sigma)
\nonumber\\ &
-\kappa (\phi^T \Delta \phi \sigma + \text{h.c.}).
\end{align}

Imposing the tadpole equations (the equations stating that the vev's are obtained at the minimum 
of the scalar potential) permits the elimination of the parameters $\mu_1^2$, $\mu_2^2$, and $\mu_3^2$ 
in favor of the vev's~\cite{Diaz:1998zg}.

When expanding around those vev's,
the real neutral fields $\chi_\sigma$, $\chi_\phi$, $\chi_\Delta$ become massive.
At the level of the Lagrangian this means that a term 
$\frac{1}{2} [\chi_\sigma \, \chi_\phi \, \chi_\Delta] M_\chi^2[\chi_\sigma \, \chi_\phi \, \chi_\Delta]^T$ 
appears, where

\begin{widetext}
\begin{eqnarray}
M_\chi^2 &=& \left[ \begin{array}{ccc} 
2\beta_1v_\sigma^2 + \frac{1}{2}\kappa v_\phi^2 \frac{v_\Delta}{v_\sigma}
& \beta_2 v_\phi v_\sigma-\kappa v_\phi v_\Delta & 
\beta_3 v_\Delta v_\sigma-\frac{1}{2}\kappa v_\phi^2 \\ 
\beta_2 v_\phi v_\sigma-\kappa v_\phi v_\Delta & 2 \lambda_1 v_\phi^2 & 
(\lambda_3+\lambda_5)v_\phi v_\Delta-\kappa v_\phi v_\sigma \\ 
\beta_3 v_\Delta v_\sigma-\frac{1}{2}\kappa v_\phi^2 & 
(\lambda_3+\lambda_5)v_\phi v_\Delta-\kappa v_\phi v_\sigma & 
2(\lambda_2+\lambda_4)v_\Delta^2 +
\frac{1}{2} \kappa v_\phi^2 \frac{v_\sigma}{v_\Delta}
 \end{array} \right].
\label{CPevenScalars}
\end{eqnarray}
\end{widetext}

By diagonalizing this matrix with 
$O_\chi M_\chi^2 O_\chi^T={\text{diag}}(m_{h_1}^2,m_{h_2}^2,m_{h_3}^2)$, one obtains
the masses of 
the neutral scalar fields $h_1$, $h_2$, and $h_3$. The fields are such that
$O_\chi [\chi_\sigma , \chi_\phi , \chi_\Delta]^T=[h_1,h_2,h_3]^T$. 
We assume that the lightest of them is the Higgs 
boson discovered in 2012 \cite{Aad:2012tfa,Chatrchyan:2012xdj}, with mass $m_{h_1}\approx 125$ GeV 
\cite{Aad:2015zhl}. 
In the present article we concentrate
on the phenomenology of the second $CP-$even Higgs boson $h_2$, the massive $CP-$odd Higgs boson $A$, and 
the charged Higgs boson $H^\pm$, in consistency with the SM-like higgs found at 
the LHC being $h_{1}$ in the ``123'' model.

The pseudoscalar fields $\varphi_\sigma$, $\varphi_\phi$, and $\varphi_\Delta$ mix due to the mass
matrix $M_\varphi^2$. The term in the Lagrangian has the form $\frac{1}{2} [\varphi_\sigma \, \varphi_\phi 
\, \varphi_\Delta] M_\varphi^2[\varphi_\sigma \, \varphi_\phi \, \varphi_\Delta]^T$ with
\begin{eqnarray}
M_\varphi^2 &=& \left[ \begin{array}{ccc} 
\frac{1}{2} \kappa v_\phi^2 \frac{v_\Delta}{v_\sigma} & 
\kappa v_\phi v_\Delta & 
\frac{1}{2}\kappa v_\phi^2 \\ 
\kappa v_\phi v_\Delta & 2 \kappa v_\Delta v_\sigma & 
\kappa v_\phi v_\sigma \\ 
\frac{1}{2} \kappa v_\phi^2 & \kappa v_\phi v_\sigma & 
\frac{1}{2} \kappa v_\phi^2 \frac{v_\sigma}{v_\Delta} \end{array}
\right].
\label{CPoddScalars}
\end{eqnarray}
By inspection, we know that there are two null eigenvalues, since two rows are linearly dependent 
of the third. The mass matrix is diagonalized by another rotation given by 
$O_\varphi M_\varphi^2 O_\varphi^T={\text{diag}}(m_{G^0}^2,m_{J}^2,m_{A}^2)$, where $G^0$ is the massless
nonphysical neutral Goldstone boson and $J$ is the massless physical Majoron. $A$ is the massive 
pseudoscalar, and $O_\varphi [\varphi_\sigma ,\, \varphi_\phi ,\, \varphi_\Delta]^T=[G^0,J,A]^T$ is
satisfied. The pseudoscalar $A$ has a mass,
\begin{equation}
m_A^2 = \frac{1}{2} \kappa \bigg( \frac{v_\sigma v_\phi^2}{v_\Delta} + \frac{v_\Delta v_\phi^2}{v_\sigma}
+ 4 v_\sigma v_\Delta \bigg).
\label{mA}
\end{equation}
A value of $\kappa$ different from zero is necessary to have a massive pseudoscalar $A$. 
For experimental reasons, we would like to take the massless Majoron as mainly singlet 
in order to comply with the well measured $Z$ boson invisible width \cite{Agashe:2014kda,Carena:2003aj}. 
Nevertheless, in the ``123'' model imposing this is unnecessary because the Majoron results mostly singlet 
as long as the triplet vev is small (see Appendix \ref{Diagonalization}).
The Majoron can acquire a small mass via different
possible mechanisms \cite{Akhmedov:1992hi}. 
In cases where this particle has a small mass, it can be a candidate for Dark Matter \cite{Berezinsky:1993fm,*Lattanzi:2008ds}.

We mention also the electrically charged scalars. The singly charged bosons $\phi^{-*}$
and $\Delta^+$ mix to form the term in the Lagrangian 
$[\phi^-\,,\Delta^{+*}]M_+^2[\phi^{-*},\Delta^{+}]^T$, with
\begin{widetext}
\begin{eqnarray}
M_+^2 &=& \left[ \begin{array}{cc} 
-\frac{1}{2} \lambda_5 v_\Delta^2 + \kappa v_\Delta v_\sigma
& \frac{1}{2\sqrt{2}}\lambda_5v_\Delta v_\phi 
- \frac{1}{\sqrt{2}} \kappa v_\phi v_\sigma \\ 
\frac{1}{2\sqrt{2}}\lambda_5v_\Delta v_\phi - \frac{1}{\sqrt{2}} \kappa v_\phi v_\sigma & 
-\frac{1}{4} \lambda_5 v_\phi^2 + \frac{1}{2} \kappa v_\phi^2 v_\sigma / v_\Delta
\end{array} \right],
\end{eqnarray}
\end{widetext}
which is diagonalized by a rotation given by $O_+M_+^2 O_+^T={\text{diag}}(m_{G^{+}}^2,m_{H^{+}}^2)$. As in 
the previous case, by inspection this mass matrix has a null eigenvalue corresponding to the charged 
Goldstone boson. The mass eigenstate fields satisfy $O_+ [\phi^{-*},\Delta^{+}]^T=[G^+,H^+]^T$.
The charged Higgs mass is,
\begin{equation}
m_{H^\pm}^2 = \frac{1}{2} \bigg( \kappa\frac{v_\sigma}{v_\Delta} - \frac{1}{2} \lambda_5 \bigg)
\Big( v_\phi^2 + 2 v_\Delta^2 \Big).
\label{mH+}
\end{equation}

Finally, the doubly charged boson $\Delta^{++}$ mass is given by
\begin{equation}
m_{++}^2 = - \lambda_4 v_\Delta^2 - \frac{1}{2} \lambda_5 v_\phi^2 +
\frac{1}{2} \kappa v_\phi^2 \frac{v_\sigma}{v_\Delta}.
\label{DoubleCharged}
\end{equation}
since it does not mix (it is purely triplet).

\section{Restrictions on the Parameter Space}
\label{ParamRestrictions}

In this Section we explain our restrictions on the model parameters. We first comment that the invisible decay 
width of the $Z$ gauge boson in our model is suppressed since the Majoron $J$ is mostly singlet ($O_\varphi^{21}\approx1$). 
We define $\Gamma_{inv}^{123}$ as the decay width of the $Z$ into undetected particles 
excluding the decay into neutrinos, $Z\rightarrow\overline{\nu}\nu$. 
Experimentally, $\Gamma_{inv}^{123}<2$ 
MeV at $95\%$ CL. \cite{Agashe:2014kda,Carena:2003aj}
and in our model there could be a contribution from 
the mode $Z\rightarrow JZ^* \rightarrow J\overline{\nu}\nu$. This contribution is automatically
suppressed because the Majoron is mainly singlet (see Appendix \ref{Diagonalization}).

Also, this model includes three $CP-$even Higgs bosons. We assume that the lightest of them is SM-like, and
therefore fits with the experimental results. That is, we assume its mass is near 125 GeV, that 
it is mainly doublet ($O_\chi^{12}\approx1$), and that its invisible decay width is negligible
\cite{Falkowski:2013dza}.
This last condition is obtained if we suppress the $h_1$ coupling to Majorons taking $|\beta_2|\le0.05$. 

The constraints we implement are:
\begin{itemize}
\item
$|O_\varphi^{21}|\ge0.95$ ($J$ mainly singlet)

\item
The $\rho$ parameter is also very well measured: $\rho = 1.00037\pm0.00023$
\cite{Agashe:2014kda}. 
In this model it is
\begin{equation}
\rho = 1 - \frac{2v_\Delta^2}{v_\phi^2+4v_\Delta^2}.
\end{equation}
This restricts the value of $v_\Delta$ to be smaller than a few GeV. Nevertheless, we consider 
$v_\Delta < 0.35$ GeV as in Ref.~\cite{Diaz:1998zg} in order to satisfy astrophysics bounds. 
\item
$m_{h_1}=125.09\pm0.24$ GeV \cite{Aad:2015zhl}.
\item
$|O_\chi^{12}|\ge0.95$ ($h_1$ mainly doublet)
\item
$|\beta_2|\le0.05$ (small $h_1$ invisible decay)
\item
$m_{H^\pm} > 80$ GeV \cite{Agashe:2014kda}.
\end{itemize}
We make a general scan where we vary all the independent parameters. We generate 
their values randomly from uniform distributions. We do our scan with positive values of
$\lambda_1$, $\beta_1$ and $\kappa$, as negative values of these parameters typically result 
in negative eigenvalues of the mass matrix in eq. (\ref{CPevenScalars}). The window for $v_2$ is reduced 
because of its dependency with the masses of the $W$ and $Z$ bosons~\cite{Akeroyd:2010eg}. 
Considering the range of $v_2$ and $v_3$, the scanned range for $\lambda_1$ is mostly fixed due to its strong 
dependency with $m_{h_1}\approx125$ GeV, and also because of the small effects of the mixings with 
other $CP-$even scalars (see eq.~(\ref{CPevenScalars})). 
Terms outside of the mass matrix diagonal are generally much smaller than those on the diagonal, making the terms in the diagonal leading almost directly to the masses of $h_1$, $h_2$ and $h_3$.
The scanned range for $\beta_2$ is forced to be small to avoid a large $h_1$ invisible decay (see Section \ref{h1decays}).

After imposing our constraints we note a clear hierarchy where $v_\sigma\gg v_\phi\gg v_\Delta$ that 
we have partially imposed: $v_\Delta$ is small in order to account for the 
measured $\rho$ parameter, and $v_\phi\approx 246$ GeV to account for 
the Higgs mass. With that, a large value for $v_\sigma$ comes naturally. 

We find a small effect from our filters in $\lambda_2$, $\lambda_3$, 
$\lambda_4$, $\lambda_5$ and $\beta_3$. We note that the value of $\kappa$ cannot be 
zero because in that case the $CP-$odd Higgs $A$ would be massless, and since it is 
mostly triplet that would contradict the measurements for the invisible decay of 
the $Z$ boson. Its value cannot be to large neither because mixing in the $CP-$even sector 
would move $h_1$ away from the mostly doublet-like scenario (a SM-like Higgs boson).
%
\begin{figure*}
\includegraphics[width=\textwidth]{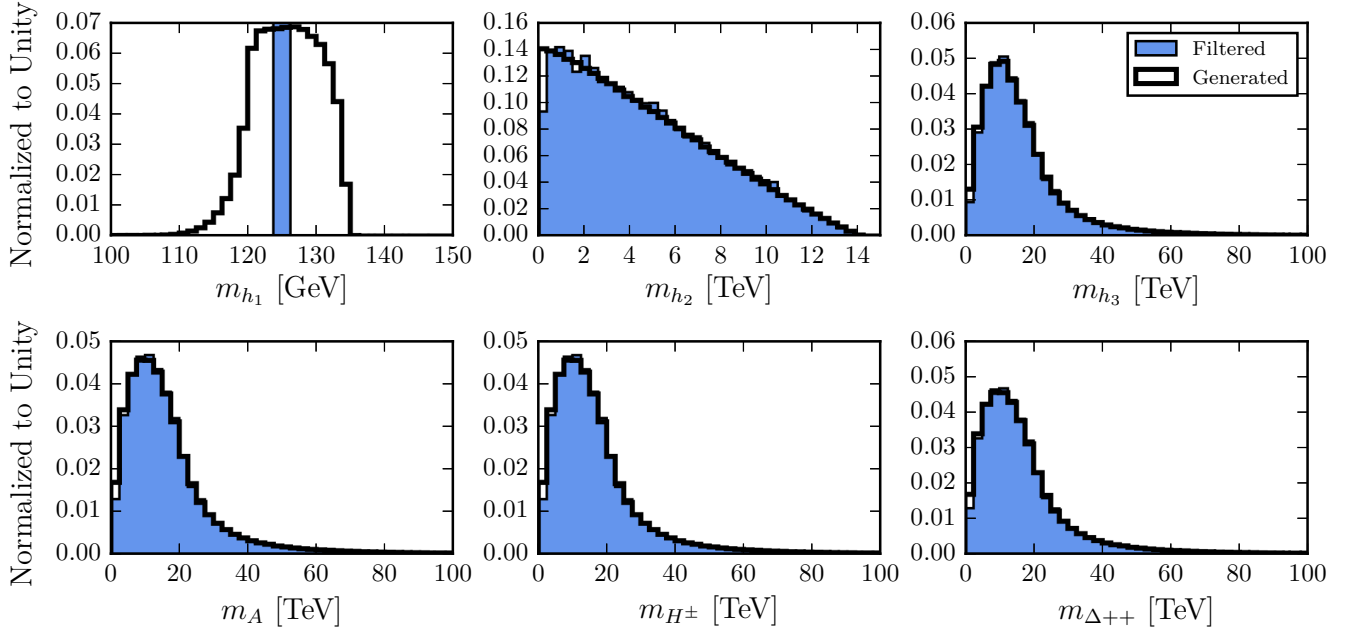}
\caption{Distribution of the physical masses in the general scan. Parameters are varied as in Table \ref{tab1}.}
\label{Masses}
\end{figure*}
%
After the scan and imposing the filters we can see the distribution of the physical masses 
in our model. This is shown in Fig.~\ref{Masses}, where the thick black line shows 
the distribution before cuts to appreciate their effect. The most distinctive 
feature is that we impose the lightest scalar mass to be $m_{h_1}\approx125$ GeV. All the other masses 
are free. The model allows for heavier scalars considering that we still have room for large parameters.

We highlight that the Majoron is massless in this model and is naturally mainly singlet, as can be 
inferred from eq.~(\ref{NeuFields123Model}), which is related to the exact diagonalization of the $CP-$odd mass matrix 
shown in Appendix \ref{Diagonalization}. Also notice that the new scalar states have the tendency
to be heavy, with extreme values for the masses obtained for high values of the parameters. The shape 
of the distributions in Fig.~(\ref{Masses}) of course depends on using a linear generation of random 
values, which highlights large masses. Anyhow, we consider this to be an argument against colliders with
small values for the centre of mass (CM) energy.

There is also an ambiguity related to the composition of the $h_2$ field: it can be mainly singlet, mainly 
triplet, or anything in between, as long as it is not mainly doublet, which is reserved for $h_1$, our 
SM-like Higgs boson. If $h_2$ is mainly triplet its mass tends to be similar to the masses of $A$,
$H^+$, and $\Delta^{++}$ (all these fields are mainly triplet). If $h_2$ is mainly singlet, 
the mass of $h_3$ tends to be equal to the masses of $A$, $H^+$, and $\Delta^{++}$, and in this case, 
a mainly-singlet $h_2$ can be lighter. The masses of $h_2$ and $h_3$ are strongly correlated with the values of $({M_\chi})^2_{11}$ and $({M_\chi})^2_{33}$ depending on which is mainly singlet or triplet. Obtaining a scenario where $h_2$ and $h_3$ are not purely singlet or triplet requires ${(M_\chi})^2_{11}$ numerically very close to $({M_\chi})^2_{33}$, making that scenario highly fine-tuned.

The splitting between the mainly triplet fields is controlled by $|\lambda_5|$. This can be algebraically 
understood starting from the hierarchy $v_\Delta\ll v_\phi,v_\sigma$ and approximating eq.~(\ref{mA}):
\begin{equation}
m_A^2 \approx \frac{1}{2} \kappa \frac{v_\sigma v_\phi^2}{v_\Delta}
\end{equation}
Using the same approximation in eqs.~(\ref{mH+}) and (\ref{DoubleCharged}), we get for the singly 
and doubly charged Higgs masses,
\begin{eqnarray}
m_{H^\pm}^2 &\approx& m_A^2 - \frac{1}{4} \lambda_5 v_\phi^2
\nonumber\\
m_{++}^2 &\approx& m_A^2 - \frac{1}{2} \lambda_5 v_\phi^2 \approx m_{H^\pm}^2 - \frac{1}{4} \lambda_5 
v_\phi^2.
\end{eqnarray}
Thus, $H^\pm$, $\Delta^{++}$ and $A$ can differ appreciably in mass as long as $|\lambda_5|$ is large.

The previous considerations motivate us to define three benchmarks, characterized by the composition 
of $h_2$ in Table \ref{tab:DefBenchs}. The parameters for each benchmark are defined in Table \ref{tab1}. 
Note that these are chosen thinking of $e^{+}e^{-}$ colliders, given the masses below 1 TeV.

\begin{table*}
\caption{Characterization of the three benchmark under study, giving the composition of $h_2$.
\label{tab:DefBenchs}}
\begin{tabular*}{\textwidth}{@{\extracolsep{\fill}} c c c c c}
\hline \hline
	Benchmark & Composition of $h_2$ & $|O_\chi^{21}|$  & $|O_\chi^{22}|$ & $|O_\chi^{23}|$\\
\hline
	B1 & mostly triplet & $1.0\times10^{-5}$ & $1.5\times 10^{-3}$  & $1.0$ \\
	B2 & mostly singlet & $1.0$ & $9.7\times 10^{-3}$ & $8.7\times10^{-4}$ \\
	B3 & mixed & $8.9\times 10^{-1}$ & $9.8\times10^{-4}$  & $4.6\times 10^{-1}$ \\
\hline \hline
\end{tabular*}
\end{table*}
\begin{table}[h]
\caption{Scanned range for the independent parameters and their values for the different benchmarks.
\label{tab1}}
\begin{tabular*}{0.5\textwidth}{@{\extracolsep{\fill}} c c c c c c}
\hline\hline
Parameter        & Scanned Range      & B1          & B2          & B3       & Units \\ \hline
$v_\sigma$       & [$0,5000$]         & $1500$      & $3300$      & $2500$   & GeV   \\
$v_\phi$         & [$245,247$]        & $246$       & $246$       & $246$    & GeV   \\
$v_\Delta$       & [$0,0.35$]         & $0.2$       & $0.2$       & $0.3$    & GeV   \\
$\lambda_1$      & [$0.127,0.15$]     & $0.13$      & $0.13$      & $0.13$   & -     \\
$\lambda_{2}$    & [$\text{-}4,4$]    & $0.1$       & $0.1$       & $0.1$    & -     \\
$\lambda_{3}$    & [$\text{-}4,4$]    & $0.1$       & $0.1$       & $0.1$    & -     \\
$\lambda_{4}$	 & [$\text{-}4,4$]    & $0.1$       & $0.1$       & $0.1$    & -     \\
$\lambda_{5}$    & [$\text{-}4,4$]    & $1.0$       & $0.5$       & $0.8$    & -     \\
$\beta_{1}$      & [$0,4$]             & $0.3$      & $0.02$      & $0.008$  & -     \\
$\beta_{2}$      & [$\text{-}0.05,0.05$] & $0.02$   & $0.005$     & $0$      & -     \\
$\beta_{3}$      & [$\text{-}4,4$]     & $0.1$      & $0.5$       & $0.6$    & -     \\
$\kappa$         & [$0,1$]             & $0.001$    & $0.0015$    & $0.0004$ & -     \\
\hline\hline
\end{tabular*}
\end{table}
We stress the fact that there is an ambiguity in the composition of $h_2$. By definition 
$h_1$ is mainly doublet. The $H^+$ and $\Delta^{++}$ fields are always mainly triplet. 
The $A$ field is also always mainly triplet because $J$ is mainly singlet. The composition 
of $h_3$ is complementary to the composition of $h_2$.

Table \ref{tab2} shows the physical masses obtained for the three 
benchmarks. In B1 $h_2$ is mainly triplet, thus it has a mass similar 
to $A$, $H^\pm$, and $\Delta^{++}$ masses, with $h_3$ heavier. In B2 $h_2$ is 
mainly singlet, thus it is $h_3$ that has a mass similar to the masses of $A$, $H^\pm$, 
and $\Delta^{++}$, with $h_2$ lighter. 
\begin{table}[h]
\caption{Physical masses in GeV for the different benchmarks.
\label{tab2}}
\begin{tabular*}{0.5\textwidth}{@{\extracolsep{\fill}} c c c c}
\hline\hline
Parameter         &  B1      & B2      & B3       \\ \hline
$m_{h_1}$         & $125$    & $125$   & $125$    \\
$m_{h_2}$         & $476$    & $660$   & $316$    \\
$m_{h_3}$         & $1162$   & $865$   & $318$    \\
$m_A$             & $476$    & $865$   & $317$    \\
$m_{H^+}$         & $460$    & $861$   & $298$    \\
$m_{\Delta^{++}}$ & $443$    & $857$   & $277$    \\
\hline\hline
\end{tabular*}
\end{table}
%
\section{Production at the LHC}
\label{LHCprod}

Here we briefly comment on the production cross-section at the LHC for the scalars $h_{2}, A$ 
and $H^{\pm}$ for our model benchmarks (which we choose thinking of $e^{+}e^{-}$ colliders). We 
implement the ``123'' HTM in {\sc{FeynRules}}~\cite{Alloul:2013bka} and 
interface the output to the {\sc{MadGraph5}}~\cite{Alwall:2014hca} event generator to 
compute production cross-sections.

When thinking of a SM-like Higgs boson (such as $h_{1}$ in our model), the main production mode 
at the LHC is gluon-gluon fusion ($ggF$),
\begin{center}
\vspace{-50pt} \hfill \\
\begin{axopicture}(150,120)(0,26) 
\Gluon(20,90)(50,60){3}{5}
\Gluon(50,30)(20,0){3}{5}
\ArrowLine(50,60)(50,30)
\ArrowLine(50,30)(70,45)
\ArrowLine(70,45)(50,60)
\DashLine(70,45)(100,45){4}
\Text(15,90)[]{$g$}
\Text(15,0)[]{$g$}
\Text(100,55)[]{$h$}
\Text(45,45)[]{$t$}
\Text(65,60)[]{$\bar{t}$}
\Text(65,30)[]{$t$}
\end{axopicture}.
\vspace{30pt} 
$ $ \hfill
\end{center}
\vspace{10pt}
This process dominates SM-like Higgs production not only because the $ht\bar{t}$ coupling is large, but 
also because the parton distribution functions indicate that it is easier to find a 
gluon inside the proton than a heavy quark or an electroweak gauge boson.

Nevertheless, this mechanism is not be efficient for a not-mainly-doublet Higgs boson (which 
is the case for $h_{2}$ and $A$ in our model benchmarks), because that Higgs couples to 
quarks very weakly. In the model studied here, the ratio of production cross-sections  
in the gluon-gluon fusion mode for $h_1$ and $h_2$ is,
\begin{equation}
\frac{\sigma(ggF,h_2)}{\sigma(ggF,h_1,m_{h_1}=m_{h_2})} = 
\left(\frac{O_\chi^{22}}{O_\chi^{12}}\right)^2 \approx (O_\chi^{22})^2.
\end{equation}
The last approximation is valid because we have $h_1$ mainly doublet (SM-like). The production cross-section 
at $\sqrt{s}=14$ TeV for $h_{2}$ reaches $5.7\times 10^{-6}$ pb in B1, $5.7\times 10^{-5}$ pb in B2 
and $3.9\times 10^{-6}$ pb in B3. For $A$ production, the above ratio is proportional 
to $(O_\varphi^{32})^2$ and we get similar numbers. The cross-section at $\sqrt{s}=14$ TeV reaches $6.8\times10^{-6}$ pb in B1, $4.0\times10^{-7}$ pb in B2 and is somewhat higher in B3, reaching $2.5\times10^{-5}$ pb. So we conclude 
that the above ratio is around $10^{-4}$ at most. This is why, if the model 
is correct, we may have not seen $h_2$ (nor $A$) at the LHC via $ggF$, as is not a dominant production mode 
since $h_2$ does not behave like a SM-like Higgs.

Other production mechanisms that can be relevant at the LHC are electroweak modes, for 
example vector boson fusion (VBF), but they also produce small cross-sections for our given 
benchmarks. When considering the sum over all VBF processes like the diagram below,
the highest cross-section at $\sqrt{s}=14$ TeV we get is $2.5\times{10^{-5}}$ pb 
for the charged Higgs production,

\begin{center}
\vspace{-50pt} \hfill \\
\begin{axopicture}(150,120)(0,26) 
\Line[arrow](10,70)(150,70)
\Line[arrow](10,10)(150,10)
\Line[dash](90,40)(150,40)
\Text(110,50)[]{$H^{+}$}
\Photon(50,70)(90,40){-3}{5}
\Photon(50,10)(90,40){-3}{5}
\Text(10,80)[]{$q$}
\Text(10,0)[]{$q$}
\Text(130,80)[]{$\bar{q}$}
\Text(130,2)[]{$\bar{q}$}
\Text(60,30)[]{$W^{+}$}
\Text(60,50)[]{$Z$}
\end{axopicture}
\vspace{30pt} 
$ $ \hfill
\end{center}
\vspace{10pt}

in B3. Production processes via quark anti-quark annihilation can also be  
relevant. In the case of $h_{2}$ production, the highest contribution comes from 
the diagram

\begin{center}
\begin{axopicture}(150,90)(0,26) 
\ArrowLine(20,60)(50,30)
\ArrowLine(50,30)(20,0)
\Photon(50,30)(90,30){-3}{5}
\DashLine(120,60)(90,30){4}
\DashLine(90,30)(120,0){4}
\Text(10,60)[]{$q$}
\Text(10,0)[]{$\bar{q}$}
\Text(70,40)[]{$W^{+}$}
\Text(100,50)[]{$H^{+}$}
\Text(130,2)[]{$h_2$}
\DashLine(120,60)(150,80){4}
\Photon(150,40)(120,60){-3}{5}
\Text(160,80)[]{$h_1$}
\Text(160,40)[]{$W^{+}$}
\end{axopicture}
\vspace{30pt} 
$ $ \hfill
\end{center}
\vspace{10pt}
for B1 and B3. The cross-section at $\sqrt{s}=14$ TeV for B1 is $4.5\times 10^{-4}$ pb.
Production of $A$ at $\sqrt{s}=14$ TeV dominates in B1 when in the above diagram we 
replace $h_{2}$ with $A$, $W^{+}$ with a $Z$, $h_{1}$ also with a $Z$ and $H^{+}$ with $h_{2}$, leading to 
the $AZZ$ final state. This gives a cross-section of $3.7\times 10^{-4}$ pb. It can go higher 
in B3 in the $AJJ$ final state, with a cross-section reaching $2.3\times 10^{-3}$ pb. Charged Higgs 
production at $\sqrt{s}=14$ TeV can reach $4.3\times 10^{-3}$ pb in B3 in 
the $H^{+}W^{-}W^{-}$ final state (replacing $W^{+}$ and $h_1$ with $W^{-}$, $H^{+}$ with $\Delta^{--}$ 
and $h_2$ with $H^{+}$ in the above diagram).

The highest cross-section found in our model benchmarks for each 
characteristic production mechanism at the LHC is summarized in Table~\ref{BiggestLHCSigma} for comparison.
\begin{table}[h]
\caption{Highest LHC production cross-section (in units of pb) found in our benchmarks for $h_2$, $A$ and $H^{\pm}$ at $\sqrt{s}=14$ TeV via the three characteristic production mechanisms: $ggF$, $VBF$ and $q\bar{q}$ annihilation.
\label{BiggestLHCSigma}}
\begin{tabular*}{0.5\textwidth}{@{\extracolsep{\fill}} cccc}
\hline\hline    
$\sigma$     &  $h_{2}$                 & $A$                     & $H^{\pm}$               \\\hline
$ggF$        &  $5.7\times10^{-5}$ (B2) & $2.5\times10^{-5}$ (B3) & $-$                     \\
$VBF$        &  $4.4\times10^{-6}$ (B3) & $2.2\times10^{-5}$ (B1) & $2.5\times10^{-5}$ (B3) \\
$q\bar{q}$   &  $4.5\times10^{-4}$ (B1) & $2.3\times10^{-3}$ (B3) & $4.3\times 10^{-3}$ (B3)\\
\hline\hline
\end{tabular*}
\end{table}

To finish, not even the HL-LHC~\cite{Vankov:2012xs,*CMS:2013xfa} will help, because it is expected 
to have a factor of $10$ increase in luminosity, and it will not compensate the smallness 
of the production cross-section.

In summary, it seems hadron colliders are not well equipped to produce the new states 
$h_2$, $A$ and $H^{\pm}$. Production for $h_2$ and $A$ via $ggF$ at the LHC is not 
efficient since these Higgs bosons are not-mainly doublet. Productions for $h_2$, $A$ and $H^{\pm}$ via VBF
can be only as large as $\sim 10^{-5}$ pb for our benchmarks. Electroweak production 
via quark anti-quark annihilation can be as high as $\sim10^{-3}$ pb. Given that our 
benchmarks are not likely to be observed at the LHC (a dedicated analysis is needed 
to confirm this), the large hadronic background at the LHC and the advantage of a cleaner collider 
environment at lepton colliders, we focus on the production for these states at future 
electron-positron colliders.

\section{Production at $e^{+}e^{-}$ colliders}
\label{eplusProd}

In order to assess the discovery potential of the model, we implement it
in {\sc{FeynRules}}~\cite{Alloul:2013bka} so we can extract 
relevant parameters and Feynman rules. We then interface the output to 
the {\sc{MadGraph5}}~\cite{Alwall:2014hca} event generator in order to 
compute production cross-sections, as we did in the previous section.

The FCC-ee machine is a hypothetical circular $e^+e^-$ collider at CERN with a high
luminosity but low energy, designed to study with precision the Higgs boson \cite{Gomez-Ceballos:2013zzn}.
We consider its highest proyected energy 350 GeV with a luminosity of $2.6 {\, \mathrm{ab}}^{-1}$, which 
was calculated by taking the $0.13 {\, \mathrm{ab}}^{-1}$ quoted in~\cite{Gomez-Ceballos:2013zzn} and assuming 
4 interaction points and 5 years of running of the experiment. 

The canonical program for the ILC \cite{Baer:2013cma} includes three 
CM energies given by 250 GeV, 500 GeV, and 1000 GeV, with integrated 
luminosities 250 fb$^{-1}$, 500 fb$^{-1}$ and 1000 fb$^{-1}$, respectively.
CLIC \cite{Abramowicz:2013tzc} has three operating CM energies: $\sqrt{s}=350$ GeV, 1.4 TeV and 3 TeV, with 
estimated luminosities $500 {\,\mathrm{fb}}^{-1}$, $1.5 {\, \mathrm{ab}}^{-1}$ and 
$2 {\, \mathrm{ab}}^{-1}$, respectively. Based on this, we compute $e^{+}e^{-}$ production 
cross-sections for $h_{2}$, $A$ and $H^{+}$ for our three benchmarks at different CM energies.

\subsection{$h_2$ Production}
\label{h2Prod}
Table \ref{tab_h2_prod} shows $h_2$ production cross-sections at $e^{+}e^{-}$ 
colliders, prospected luminosities and CM energies for the FCC-ee, ILC and CLIC colliders.
The cross-sections are calculated by summing all
$e^+e^-\rightarrow h_2 X Y$ 3-body production modes, plus the 2-body production modes 
$e^+e^-\rightarrow h_2 X$, where $X$ is a particle that does not decay.
\begin{table*}
\caption{Production cross-section (in units of ab) for $h_2$ at an $e^+e^-$ collider for projected energies
in the 3 benchmarks. Estimated luminosities are also given in units of ab${}^{-1}$.
\label{tab_h2_prod}}
\begin{tabular*}{\textwidth}{@{\extracolsep{\fill}} ccccccc}
\hline\hline
$\sqrt{s}$ [TeV] & ${\cal L}_{FCCee}$ & ${\cal L}_{ILC}$ & ${\cal L}_{CLIC}$ & B1: $\sigma$ & B2: $\sigma$ & B3: $\sigma$ \\\hline
$0.250$&-    &$0.25$&$-$  &$0$               &$0$               &$0$                \\
$0.350$&$2.6$&$-$   &$0.5$&$0$               &$0$               &$1.7\times10^{-5}$ \\
$0.500$&-    &$0.5$ &$-$  &$3.1\times10^{-6}$&$0$               &$2.5\times10^{-2}$ \\
$1.0$  &-    &$1$   &$-$  &$1.4\times10^{3}$ &$0.9$             &$3.7\times10^{3}$  \\
$1.4$  &-    &$-$   &$1.5$&$1.1\times10^{4}$ &$3.6$             &$4.1\times10^{3}$  \\
$3$    &-    &$-$   &$2$  &$6.1\times10^{3}$ &$3.5\times10^{-2}$&$2.0\times10^{3}$  \\
\hline\hline
\end{tabular*}
\end{table*}
The production cross-sections shown in Table~\ref{tab_h2_prod} are dominated by 
the 2-body production process (or mode) $e^+e^-\rightarrow h_2A$ and by
3-body production processes as follows. 
In B1 the process $e^+e^-\rightarrow h_2t\bar{t}$ is the most important one. In B2 the dominating process 
is $e^+e^-\rightarrow h_2Ah_1$. In B3 the process $e^+e^-\rightarrow h_2Zh_1$ is the dominant one. All of 
them are enhanced when a second heavy particle is also on-shell. 
%
\begin{figure*}
\includegraphics[width=\textwidth]{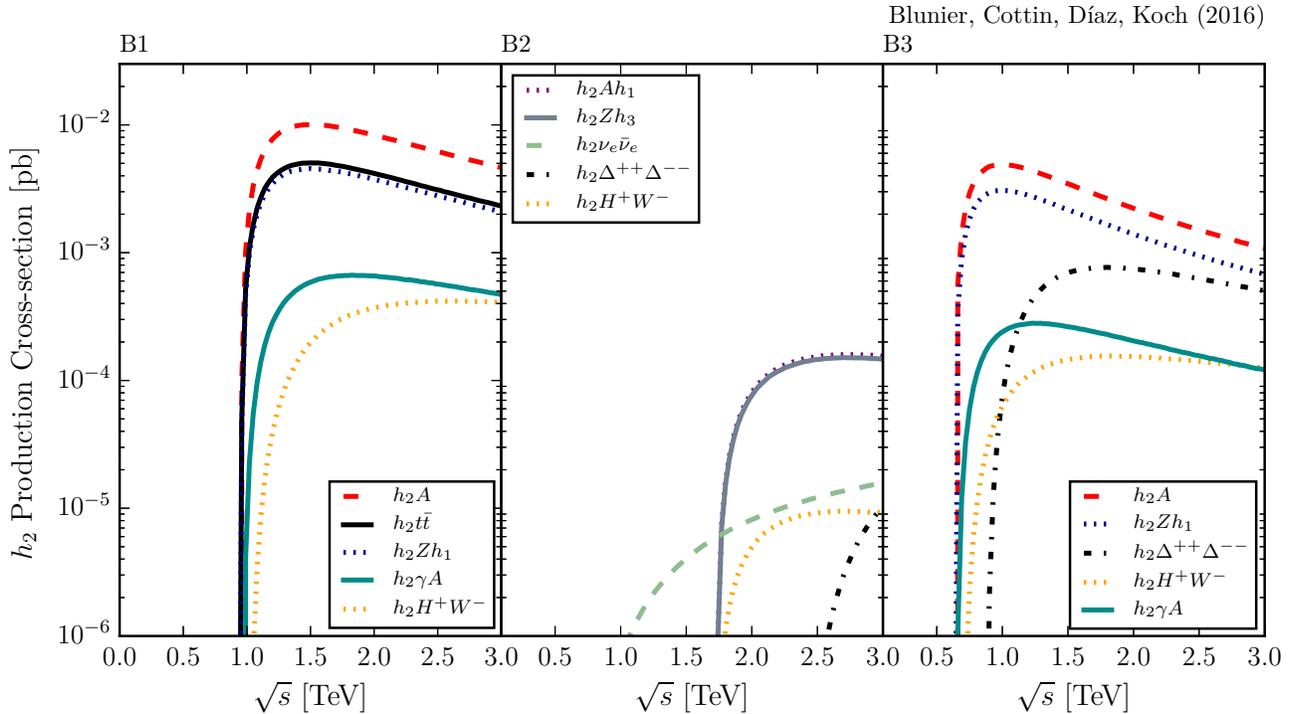}
\caption{Production modes for $h_2$ at an $e^{+}e^{-}$ collider in the 3 benchmarks. The legend shows the final 
state after the $e^{+}e^{-}$ collision.}
\label{h2_prod}
\end{figure*}
%
We show in Fig.~\ref{h2_prod} the main $h_2$ production modes for all 3 benchmarks. 
In B1 (left frame) this particle is potentially observed at CLIC only 
when the $A$ scalar is also on-shell. Thus, the main 2-body production mode is the 
so-called associated production,
\begin{center}
\vspace{-50pt} \hfill \\
\begin{picture}(150,90)(0,26) 
\ArrowLine(20,60)(50,30)
\ArrowLine(50,30)(20,0)
\Photon(50,30)(90,30){-3}{5}
\DashLine(120,60)(90,30){4}
\DashLine(90,30)(120,0){4}
\Text(10,60)[]{$e^-$}
\Text(10,0)[]{$e^+$}
\Text(70,40)[]{$Z$}
\Text(130,60)[]{$A$}
\Text(130,2)[]{$h_2$}
\end{picture}
\vspace{30pt} 
$ $ \hfill
\end{center}
\vspace{10pt}
defined when $h_2$ is produced together with an $A$. The coupling $ZAh_2$ is given in Appendix 
\ref{FeynmanRules}. Since $A$ is mainly triplet, $O^{33}_\varphi$ is of order 1. In addition, in B1 $h_2$ is mainly triplet, 
thus $O^{23}_\chi$ is also of order 1. Therefore, the whole coupling $ZAh_2$ is not suppressed with respect 
to the gauge coupling $g$.

The most important 3-body production modes in B1 are also displayed in the left frame of 
Fig.~\ref{h2_prod}. 
The main production process is $h_2t\overline{t}$ when $A$ is on-shell. Diagramatically it looks like,
\begin{widetext}
\begin{center}
\vspace{-50pt} \hfill \\
\begin{axopicture}(350,90)(0,26) 
\ArrowLine(20,60)(50,30)
\ArrowLine(50,30)(20,0)
\Photon(50,30)(90,30){-3}{5}
\ArrowLine(120,0)(90,30)
\ArrowLine(90,30)(105,45)
\ArrowLine(105,45)(120,60)
\DashLine(105,45)(120,30){4}
\Text(10,60)[]{$e^-$}
\Text(10,0)[]{$e^+$}
\Text(70,40)[]{$Z,\gamma$}
\Text(128,60)[]{$t$}
\Text(128,2)[]{$\overline{t}$}
\Text(130,30)[]{$h_2$}
\Text(150,30)[]{$+$}
\ArrowLine(180,60)(210,30)
\ArrowLine(210,30)(180,0)
\Photon(210,30)(250,30){-3}{5}
\DashLine(280,60)(250,30){4}
\DashLine(250,30)(280,0){4}
\Text(170,60)[]{$e^-$}
\Text(170,0)[]{$e^+$}
\Text(230,40)[]{$Z$}
\Text(255,50)[]{$A,J$}
\Text(290,2)[]{$h_2$}
\ArrowLine(280,60)(310,80)
\ArrowLine(310,40)(280,60)
\Text(315,80)[]{$t$}
\Text(315,40)[]{$\overline{t}$}
\end{axopicture}
\vspace{30pt} 
$ $ \hfill
\end{center}
\end{widetext}
\vspace{10pt}
plus a similar graph with $h_2$ emitted from the anti-quark and another graph with the $A$ boson
being replaced by a $Z$ boson.
This production process is enhanced when the $A$ scalar boson is on-shell, 
$e^+e^-\rightarrow h_2 A\rightarrow h_2 t \overline{t}$, corroborated by the fact that 
$B(A\rightarrow  t \overline{t})=0.5$ is large for B1, as shown in Table \ref{tab6}.

In the central frame of Fig.~\ref{h2_prod} we see B2. In this case, production cross-sections 
are systematically smaller because in this benchmark
$h_2$ is mainly singlet and couplings to gauge bosons are smaller. Also the main production modes are 
different. The process $e^+e^-\rightarrow h_2 t \overline{t}$ is no longer efficient, with a cross-section
of the order of $10^{-8}$ pb and outside of the plot. The reason is that the coupling $Z h_2 A$ is
small when $h_2$ is mainly singlet. The main production mode for B2 is 
$e^+e^-\rightarrow h_2 A h_1$, with Feynman diagrams for the sub-processes given by,
\begin{widetext}
\begin{center}
\vspace{-50pt} \hfill \\
\begin{axopicture}(380,100)(0,26) 
\ArrowLine(20,60)(50,30)
\ArrowLine(50,30)(20,0)
\Photon(50,30)(90,30){-3}{5}
\DashLine(120,60)(90,30){4}
\DashLine(90,30)(120,0){4}
\Text(10,60)[]{$e^-$}
\Text(10,0)[]{$e^+$}
\Text(70,40)[]{$Z$}
\Text(100,50)[]{$h_i$}
\Text(130,2)[]{$A$}
\DashLine(120,60)(150,80){4}
\DashLine(150,40)(120,60){4}
\Text(160,80)[]{$h_2$}
\Text(160,40)[]{$h_1$}
\Text(180,30)[]{$+$}
\ArrowLine(210,60)(240,30)
\ArrowLine(240,30)(210,0)
\Photon(240,30)(280,30){-3}{5}
\DashLine(310,60)(280,30){4}
\DashLine(280,30)(310,0){4}
\Text(200,60)[]{$e^-$}
\Text(200,0)[]{$e^+$}
\Text(260,40)[]{$Z$}
\Text(285,50)[]{$A,J$}
\Text(325,2)[]{$h_1$}
\DashLine(310,60)(340,80){4}
\DashLine(340,40)(310,60){4}
\Text(350,80)[]{$h_2$}
\Text(350,40)[]{$A$}
\end{axopicture}
\vspace{30pt} 
$ $ \hfill
\end{center}
\end{widetext}
\vspace{10pt}
plus Feynman diagrams 
where in the last sub-process we replace $(A,J)$ by $Z$ and/or interchange $h_1$ with $h_2$. This mode is 
enhanced when $h_3$ is on-shell, since in B2 $h_3$ is mainly triplet and the coupling $Z A h_3$
is large resulting in $e^+e^-\rightarrow h_3 A\rightarrow h_2 h_1 A$.

B3 is an intermediate situation. Even in this case, $h_2$ production cross-sections are potentially observable 
when $A$ is also on-shell. The production cross-section $e^+e^-\rightarrow h_2 A$ is smaller than in B1, 
but still large. The main 3-body production mode in this case is $e^+e^-\rightarrow h_2 Z h_1$, with
sub-processes given by,
\begin{widetext}
\begin{center}
\vspace{-50pt} \hfill \\
\begin{axopicture}(380,100)(0,26) 
\ArrowLine(20,60)(50,30)
\ArrowLine(50,30)(20,0)
\Photon(50,30)(90,30){-3}{5}
\DashLine(120,60)(90,30){4}
\Photon(90,30)(120,0){-3}{5}
\Text(10,60)[]{$e^-$}
\Text(10,0)[]{$e^+$}
\Text(70,40)[]{$Z$}
\Text(100,50)[]{$h_i$}
\Text(130,2)[]{$Z$}
\DashLine(120,60)(150,80){4}
\DashLine(150,40)(120,60){4}
\Text(160,80)[]{$h_1$}
\Text(160,40)[]{$h_2$}
\Text(180,30)[]{$+$}
\ArrowLine(210,60)(240,30)
\ArrowLine(240,30)(210,0)
\Photon(240,30)(280,30){-3}{5}
\DashLine(310,60)(280,30){4}
\DashLine(280,30)(310,0){4}
\Text(200,60)[]{$e^-$}
\Text(200,0)[]{$e^+$}
\Text(260,40)[]{$Z$}
\Text(285,50)[]{$A,J$}
\Text(330,2)[]{$h_1,h_2$}
\DashLine(310,60)(340,80){4}
\Photon(340,40)(310,60){-3}{5}
\Text(360,80)[]{$h_2,h_1$}
\Text(350,40)[]{$Z$}
\end{axopicture}
\vspace{30pt} 
$ $ \hfill
\end{center}
\end{widetext}
\vspace{10pt}
where $i=1,2,3$, and missing are a graph with the $CP-$odd scalar replaced by a $Z$ and one formed with 
a $ZZh_1h_2$ quartic coupling.
This production mode is enhanced when the $A$ boson is on-shell, 
$e^+e^-\rightarrow h_2 A\rightarrow h_2 h_1 Z$, with a branching fraction  
$B(A\rightarrow  h_1 Z)=0.9$ as shown in Table \ref{tab6}.
%
\begin{figure*}
\includegraphics[width=\textwidth]{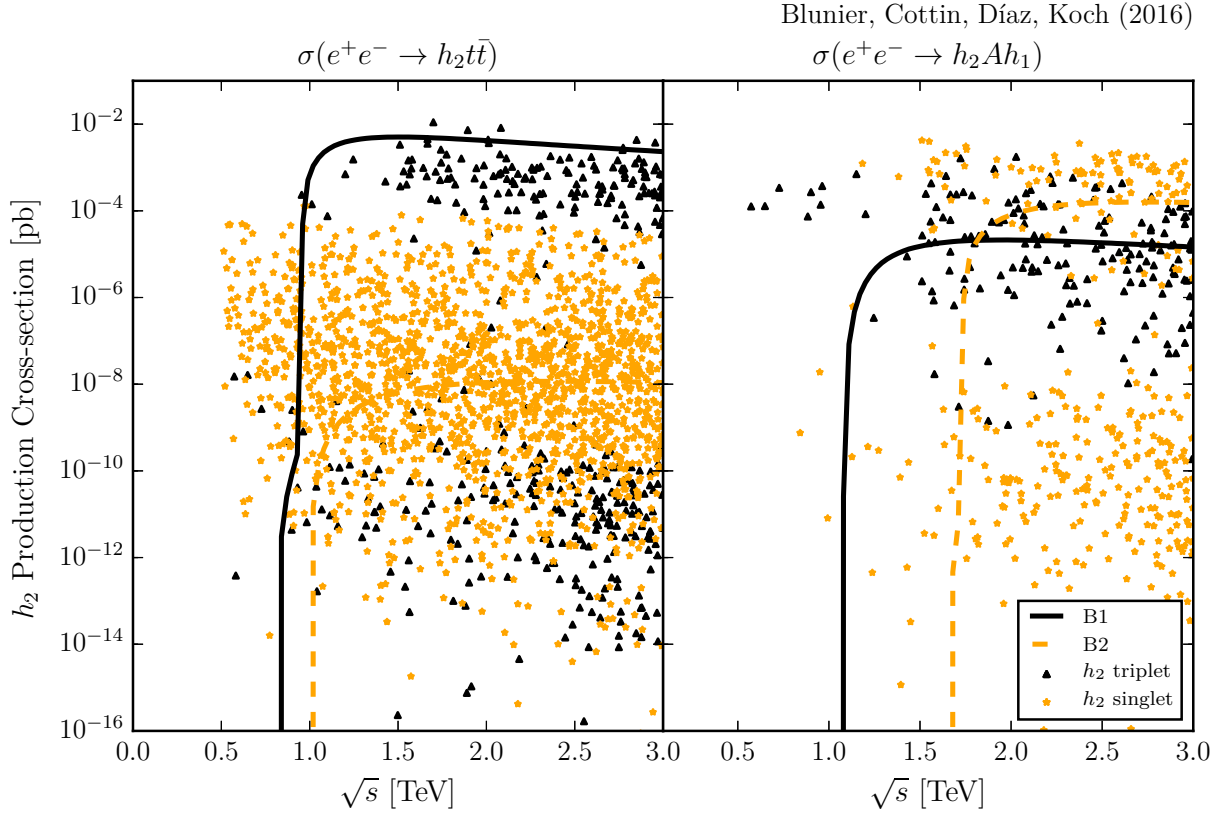}
\caption{
Production modes $e^+e^-\rightarrow h_2 t\bar{t}$ and $e^+e^-\rightarrow h_2 h_1 A$.}
\label{cloud_h2_prod}
\end{figure*}
%

Fig.~\ref{cloud_h2_prod} shows a scan for the production mode 
$e^+e^-\rightarrow h_2 t \overline{t}$ (left frame) and $e^+e^-\rightarrow h_2 h_1 A$ (right frame), two 
of the important 3-body $h_2$ production modes. In the case of $e^+e^-\rightarrow h_2 t \overline{t}$, the
production cross-section reaches up to 0.01 pb. The largest cross-sections are seen when $h_2$ is mainly 
triplet (black triangular points), with a typical value between 0.001 and 0.01 pb. B1 is shown as a black solid curve.
The value of the cross-section drops when $h_2$ is mainly singlet (orange star points), with values 
typically smaller than $10^{-4}$ pb. 
This is because a singlet does not couple to the $Z$ gauge boson. The chosen B2 lies within the cloud of points. 
The case where $h_2$ is mixed is much more rare and no point has been generated in this scenario due to its fine-tuned character. 

The case of $e^+e^-\rightarrow h_2 A h_1$ is shown in the right frame of Fig.~\ref{cloud_h2_prod}.
This is the main process in B2, where $h_2$ is mainly singlet (orange star points). In this case, cross-sections can reach up to $10^{-3}$ pb, but can also be as low as $10^{-14}$ pb,
depending on whether $h_3$ is on-shell or not. In the case where $h_2$ is
mainly triplet (black triangular points) the cross-section is more restricted. It can vary between $10^{-3}$
and $10^{-8}$ pb and B1 is a very typical case. Cross-sections are larger when an intermediate heavy
scalar is also on-shell.

Notice that the popular modes for the production of a SM-like Higgs boson in a $e^+e^-$
collider, known collectively as vector boson fusion,  
$e^+e^-\rightarrow h_2 e^+ e^-$ (fusion of two $Z$ bosons) or $e^+e^-\rightarrow h_2 \nu_e \bar{\nu}_e$
(fusion of two $W$ bosons) do not work in our case because the $h_2$ couplings to vector bosons are
suppressed by the triplet vev $v_\Delta$. In addition, most of the charged leptons go through the 
beam pipe, thus $\sigma(e^+e^-\rightarrow h_2 e^+ e^-)$ is further penalized when a cut on 
the charged lepton pseudo-rapidity is imposed. 
We use {\sc MadGraph5} default cuts, which impose that the absolute value of the charged lepton pseudo-ratidity is smaller than 2.5.

\subsection{$A$ Production}

Table \ref{tab_A_prod} shows $A$ production at $e^{+}e^{-}$ colliders, prospected luminosities 
and CM energies for the FCC-ee, ILC and CLIC colliders.
\begin{table*}
\caption{Production cross-section (in units of ab) for $A$ at an $e^+e^-$ collider for projected energies
in the 3 benchmarks. Estimated luminosities are also given in units of ab${}^{-1}$. 
\label{tab_A_prod}}
\begin{tabular*}{\textwidth}{@{\extracolsep{\fill}} ccccccc}
\hline\hline
$\sqrt{s}$ [TeV] & ${\cal L}_{FCCee}$ & ${\cal L}_{ILC}$ & ${\cal L}_{CLIC}$ & B1: $\sigma$ & B2: $\sigma$ & B3: $\sigma$ \\\hline
$0.250$&-    &$0.25$&$-$  &$0$                &$0$               &$0$                 \\
$0.350$&$2.6$&$-$   &$0.5$&$0$                &$0$               &$1.4\times10^{-10}$ \\
$0.500$&-    &$0.5$ &$-$  &$1.5\times10^{-12}$&$0$               &$1.5\times10^{-2}$  \\
$1.0$  &-    &$1$   &$-$  &$1.4\times10^{3}$  &$2.2\times10^{-5}$&$2.5\times10^{4}$   \\
$1.4$  &-    &$-$   &$1.5$&$1.1\times10^{4}$  &$3.5\times10^{-3}$&$2.1\times10^{4}$   \\
$3$    &-    &$-$   &$2$  &$6.2\times10^{3}$  &$3.6\times10^{3}$ &$7.5\times10^{3}$   \\
\hline\hline
\end{tabular*}
\end{table*}
The cross-sections are calculated in the same manner explained before.
In B1 and B2 the dominating process is $e^+e^-\rightarrow AZZ$, and in B3 the dominating process is 
$e^+e^-\rightarrow AJJ$, and all of them are enhanced when a second heavy particle is also on-shell.

\begin{figure*}
\includegraphics[width=\textwidth]{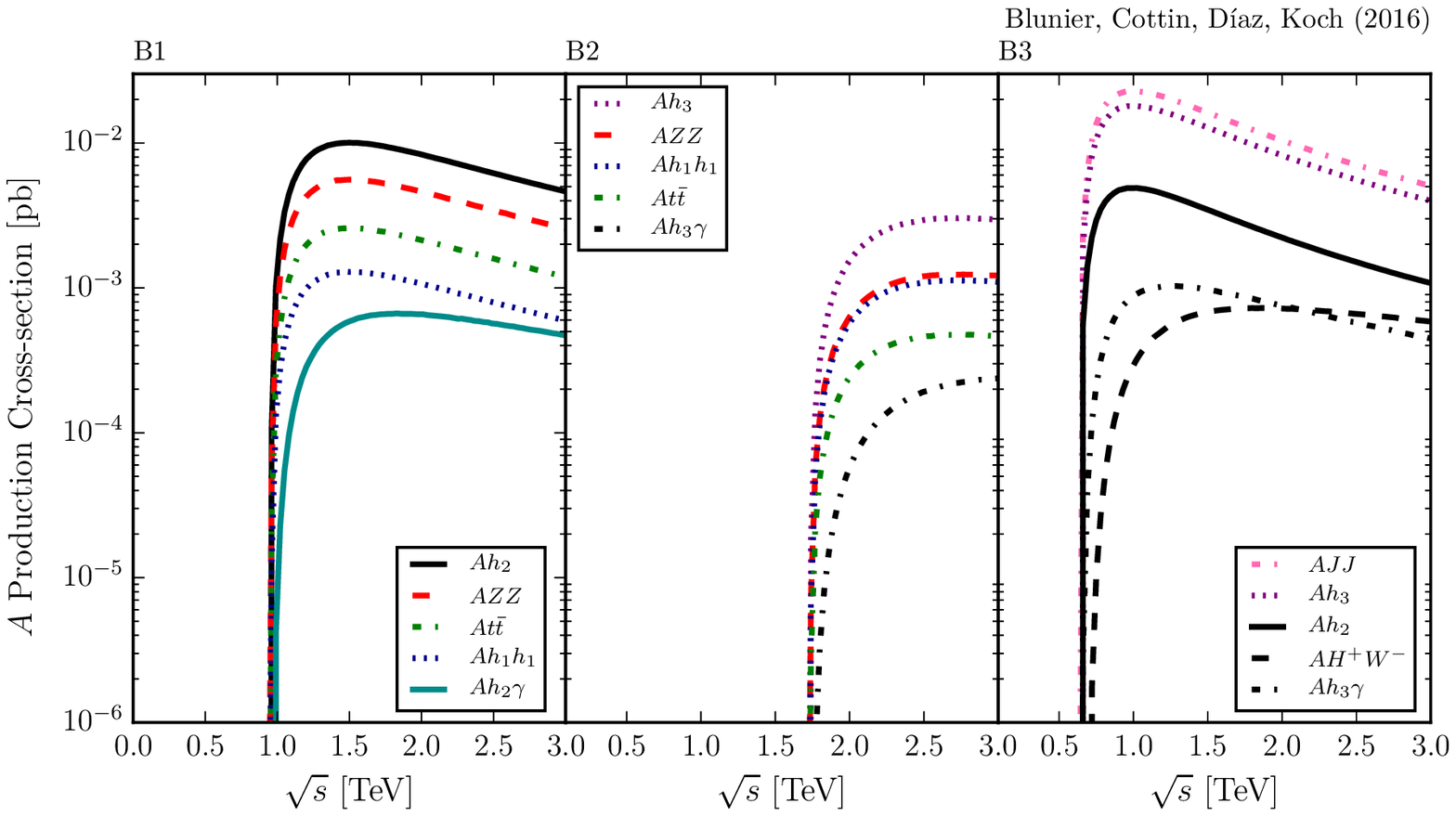}
\caption{ Production modes for $A$ at an $e^{+}e^{-}$ collider in all 3 benchmarks. The legend shows the final 
state after the $e^{+}e^{-}$ collision.}
\label{A_prod}
\end{figure*}

Fig.~\ref{A_prod} shows the production cross-sections for an $A$ boson. In B1 (left 
frame) $A$ is potentially observable at CLIC when produced in association with an $h_2$. In this case
the mode $e^+e^-\rightarrow Ah_1$ is suppressed because $O_\varphi^{32}$ and $O_\chi^{13}$ are both 
small (see Feynman rule in Appendix \ref{FeynmanRules}), thus the coupling $h_1 AZ$ itself is suppressed 
with respect to $g$.
%
%
Three body production modes are also in Fig.~\ref{A_prod}. The dominant 3-body production mode in 
B1 is $e^+e^- \rightarrow A Z Z$, represented by the Feynman diagrams,
\begin{widetext}
\begin{center}
\vspace{-50pt} \hfill \\
\begin{axopicture}(380,100)(0,26) 
\ArrowLine(20,60)(50,30)
\ArrowLine(50,30)(20,0)
\Photon(50,30)(90,30){-3}{5}
\DashLine(120,60)(90,30){4}
\Photon(90,30)(120,0){-3}{5}
\Text(10,60)[]{$e^-$}
\Text(10,0)[]{$e^+$}
\Text(70,40)[]{$Z$}
\Text(100,50)[]{$h_i$}
\Text(130,2)[]{$Z$}
\DashLine(120,60)(150,80){4}
\Photon(150,40)(120,60){-3}{5}
\Text(160,80)[]{$A$}
\Text(160,40)[]{$Z$}
\Text(180,30)[]{$+$}
\ArrowLine(210,60)(240,30)
\ArrowLine(240,30)(210,0)
\Photon(240,30)(280,30){-3}{5}
\DashLine(310,60)(280,30){4}
\DashLine(280,30)(310,0){4}
\Text(200,60)[]{$e^-$}
\Text(200,0)[]{$e^+$}
\Text(260,40)[]{$Z$}
\Text(290,50)[]{$h_i$}
\Text(320,2)[]{$A$}
\Photon(310,60)(340,80){3}{5}
\Photon(340,40)(310,60){-3}{5}
\Text(350,80)[]{$Z$}
\Text(350,40)[]{$Z$}
\end{axopicture}
\vspace{30pt} 
$ $ \hfill
\end{center}
\end{widetext}
\vspace{30pt}
It is enhanced when $h_2$ is on-shell, with a branching fraction $B(h_2\rightarrow ZZ)=0.6$, as indicated
in Table \ref{tab5}. As explained later in the decay Section, the coupling $h_2 ZZ$ is large if 
$h_2$ is mainly triplet (B1). 

In B2 the $CP-$even Higgs boson created in association with $A$ is no longer $h_2$ but $h_3$.
If $h_2$ is mainly singlet, $h_3$ is mainly triplet, and the coupling $ZAh_3$ is not suppressed. This 
is confirmed in the central frame of Fig.~\ref{A_prod} where we have B2. The most important 
2-body production mode is precisely $e^+e^- \rightarrow A h_3$, represented by the Feynman diagram
\begin{center}
\vspace{-50pt} \hfill \\
\begin{axopicture}(150,90)(0,26) 
\ArrowLine(20,60)(50,30)
\ArrowLine(50,30)(20,0)
\Photon(50,30)(90,30){-3}{5}
\DashLine(120,60)(90,30){4}
\DashLine(90,30)(120,0){4}
\Text(10,60)[]{$e^-$}
\Text(10,0)[]{$e^+$}
\Text(70,40)[]{$Z$}
\Text(130,60)[]{$A$}
\Text(130,2)[]{$h_3$}
\end{axopicture}
\vspace{30pt} 
$ $ \hfill
\end{center}
\vspace{10pt}
Also in the central frame of Fig.~\ref{A_prod} we see the main 3-body $A$ production modes. 
The most important one is again $e^+e^- \rightarrow A Z Z$, and it is enhanced 
when $h_3$ is on-shell. 

B3 is an intermediate case, and we can see in the right frame of Fig.~\ref{A_prod} that the
two 2-body production modes $e^+e^- \rightarrow A h_2$ and $e^+e^- \rightarrow A h_3$ are important 
since both $h_2$ and $h_3$ have a large triplet component.
Among the 3-body production modes, the largest one is $e^+e^- \rightarrow A JJ$,
\begin{widetext}
\begin{center}
\vspace{-50pt} \hfill \\
\begin{picture}(380,100)(0,26) 
\ArrowLine(20,60)(50,30)
\ArrowLine(50,30)(20,0)
\Photon(50,30)(90,30){-3}{5}
\DashLine(120,60)(90,30){4}
\DashLine(90,30)(120,0){4}
\Text(10,60)[]{$e^-$}
\Text(10,0)[]{$e^+$}
\Text(70,40)[]{$Z$}
\Text(100,50)[]{$h_i$}
\Text(130,2)[]{$A$}
\DashLine(120,60)(150,80){4}
\DashLine(150,40)(120,60){4}
\Text(155,80)[]{$J$}
\Text(155,40)[]{$J$}
\Text(180,30)[]{$+$}
\ArrowLine(210,60)(240,30)
\ArrowLine(240,30)(210,0)
\Photon(240,30)(280,30){-3}{5}
\DashLine(310,60)(280,30){4}
\DashLine(280,30)(310,0){4}
\Text(200,60)[]{$e^-$}
\Text(200,0)[]{$e^+$}
\Text(260,40)[]{$Z$}
\Text(289,50)[]{$h_i$}
\Text(320,2)[]{$J$}
\DashLine(310,60)(340,80){4}
\DashLine(340,40)(310,60){4}
\Text(350,80)[]{$J$}
\Text(350,40)[]{$A$}
\end{picture}
\vspace{30pt} 
$ $ \hfill
\end{center}
\end{widetext}
\vspace{10pt}
and it is enhanced when $h_2$ and $h_3$ are on-shell.

%
\begin{figure*}
\includegraphics[width=\textwidth]{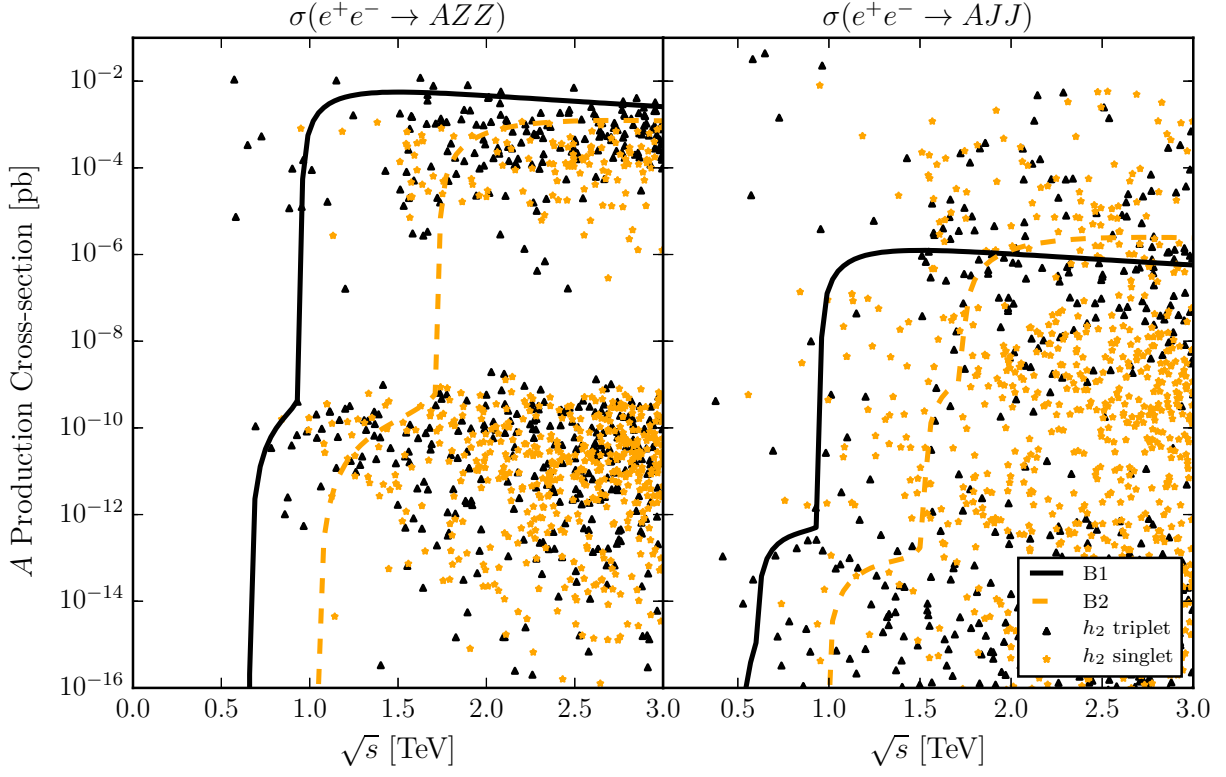}
\caption{
Production modes $e^+e^-\rightarrow A ZZ$ and $e^+e^-\rightarrow A JJ$.}
\label{cloud_A_prod}
\end{figure*}
%
Fig.~\ref{cloud_A_prod} shows scans for the process $e^+e^-\rightarrow AZZ$ (left frame), important 
for B1 and B2, and the process $e^+e^-\rightarrow AJJ$ (right frame), important in B3. 
In the first case, the production cross-section is increased when $h_2$ is also on-shell, as explained 
before. The cross-section is not larger than 0.01 pb, and B1 is not far below from that value.
In the last process a triple scalar coupling is important, and the exact values of the parameters in the potential are crucial.
In this case, B3 is characterized by a large value of $\beta_3$ which increases the coupling $h_3JJ$.
As before, in Fig.~\ref{cloud_A_prod} we include the curves corresponding to each benchmark to 
facilitate comparisons. 

\subsection{$H^+$ Production}
Table \ref{tab_Hp_prod} shows $H^+$ production cross-sections  
at $e^{+}e^{-}$ colliders, prospected luminosities and CM energies for the FCC-ee, ILC and CLIC colliders.
\begin{table*}
\caption{Production cross-section (in units of ab) for $H^+$ at an $e^+e^-$ collider for projected energies
in the 3 benchmarks. Estimated luminosities are also given in units of ab${}^{-1}$.
\label{tab_Hp_prod}}
\begin{tabular*}{\textwidth}{@{\extracolsep{\fill}} ccccccc}
\hline\hline
$\sqrt{s}$ [TeV] & ${\cal L}_{FCCee}$ & ${\cal L}_{ILC}$ & ${\cal L}_{CLIC}$ & B1: $\sigma$ & B2: $\sigma$ & B3: $\sigma$ \\\hline
$0.250$&-    &$0.25$&$-$  &$0$                &$0$                &$0$               \\
$0.350$&$2.6$&$-$   &$0.5$&$0$                &$0$                &$5.8\times10^{-3}$\\
$0.500$&-    &$0.5$ &$-$  &$1.9\times10^{-4}$ &$0$                &$0.5$            \\
$1.0$  &-    &$1$   &$-$  &$1.6\times10^{3}$  &$4.1\times10^{-3}$ &$1.7\times10^{4}$ \\
$1.4$  &-    &$-$   &$1.5$&$7.0\times10^{3}$  &$3.5\times10^{-2}$ &$1.5\times10^{4}$ \\
$3$    &-    &$-$   &$2$  &$5.0\times10^{3}$  &$2.4\times10^{3}$  &$6.6\times10^{3}$ \\
\hline\hline
\end{tabular*}
\end{table*}
Besides the 2-body production cross-section for $e^+e^-\rightarrow H^+H^-$, in
B1 and B2 the 3-body process $e^+e^-\rightarrow H^+h_1W^-$ dominates. In B3 the process 
$e^+e^-\rightarrow H^+W^+\Delta^{--}$ dominates. The last case presents a high interest, as the 
doubly charged Higgs boson gives us an independent window to study neutrinos.

\begin{figure*}
\includegraphics[width=\textwidth]{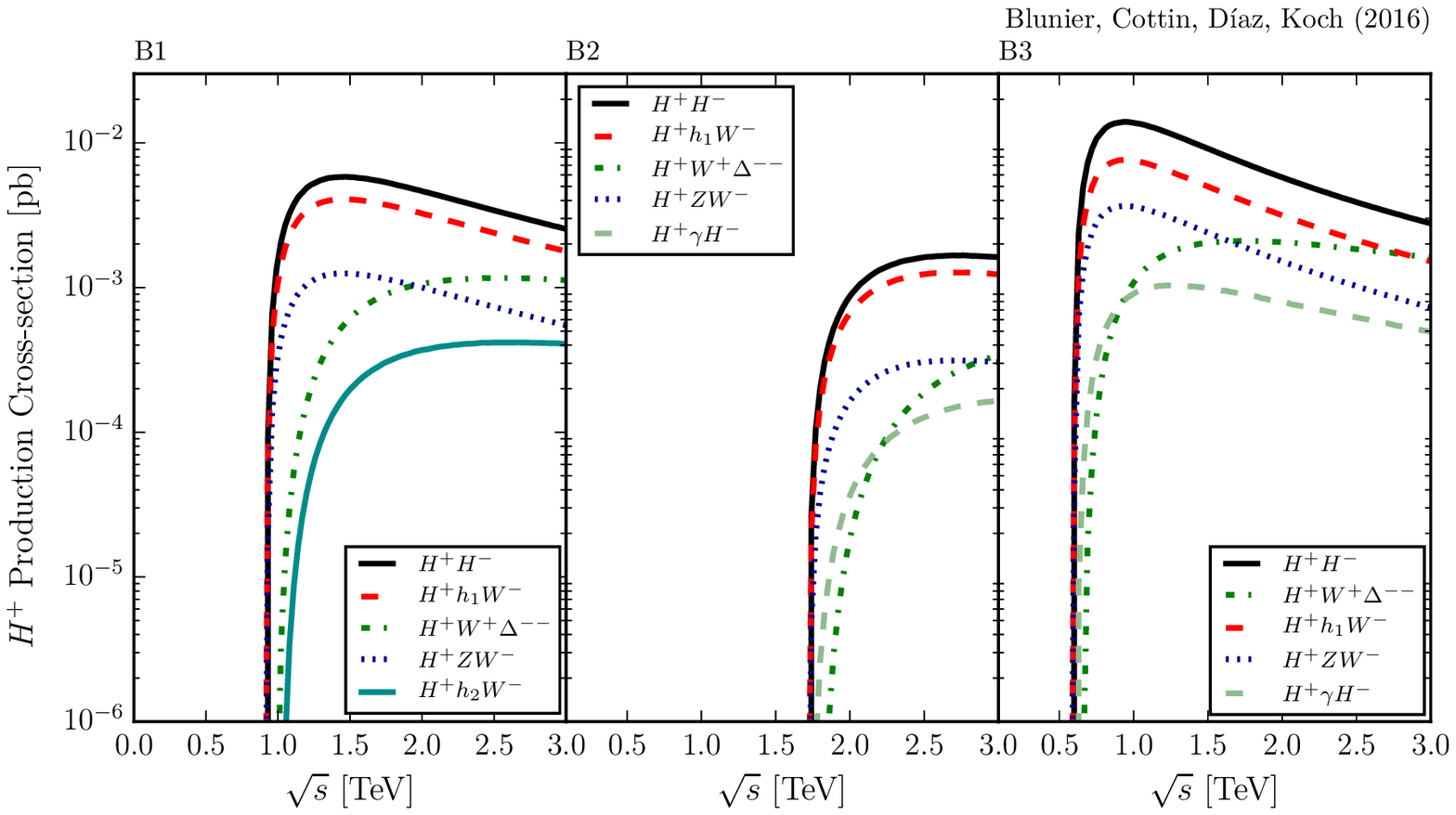}
\caption{Production modes for $H^+$ at an $e^{+}e^{-}$ collider in all 3 benchmarks. The legend shows the final state 
after the $e^{+}e^{-}$ collision.}
\label{Hp_prod}
\end{figure*}

Fig.~\ref{Hp_prod} shows the 2-body and 3-body production of an $H^+$ boson. The charged Higgs 
boson is potentially observable at CLIC when produced in association with another $H^-$, represented by the graph,
\begin{center}
\vspace{-50pt} \hfill \\
\begin{axopicture}(150,90)(0,26) 
\ArrowLine(20,60)(50,30)
\ArrowLine(50,30)(20,0)
\Photon(50,30)(90,30){-3}{5}
\DashLine(120,60)(90,30){4}
\DashLine(90,30)(120,0){4}
\Text(10,60)[]{$e^-$}
\Text(10,0)[]{$e^+$}
\Text(70,40)[]{$Z,\gamma$}
\Text(130,60)[]{$H^+$}
\Text(130,2)[]{$H^-$}
\end{axopicture}
\vspace{30pt} 
$ $. \hfill
\end{center}
\vspace{30pt}
The couplings $H^+H^-\gamma$ and $H^+H^-Z$ are both of the order of electroweak couplings, as can 
be seen in Appendix \ref{FeynmanRules}.
%
%
%
%
Among the 3-body modes, in B1 and B2 the main production mode is
$e^+e^-\rightarrow H^+h_1W^-$, represented by the sub-processes,
\begin{widetext}
\begin{center}
\vspace{-50pt} \hfill \\
\begin{axopicture}(380,100)(0,26) 
\ArrowLine(20,60)(50,30)
\ArrowLine(50,30)(20,0)
\Photon(50,30)(90,30){-3}{5}
\DashLine(120,60)(90,30){4}
\DashLine(90,30)(120,0){4}
\Text(10,60)[]{$e^-$}
\Text(10,0)[]{$e^+$}
\Text(70,40)[]{$Z,\gamma$}
\Text(100,50)[]{$H^-$}
\Text(130,2)[]{$H^+$}
\DashLine(120,60)(150,80){4}
\Photon(150,40)(120,60){-3}{5}
\Text(160,80)[]{$h_1$}
\Text(165,40)[]{$W^-$}
\Text(180,30)[]{$+$}
\ArrowLine(210,60)(240,30)
\ArrowLine(240,30)(210,0)
\Photon(240,30)(280,30){-3}{5}
\DashLine(310,60)(280,30){4}
\DashLine(280,30)(310,0){4}
\Text(200,60)[]{$e^-$}
\Text(200,0)[]{$e^+$}
\Text(260,40)[]{$Z$}
\Text(289,50)[]{$A,J$}
\Text(320,2)[]{$h_1$}
\DashLine(310,60)(340,80){4}
\Photon(340,40)(310,60){-3}{5}
\Text(350,80)[]{$H^+$}
\Text(355,40)[]{$W^-$}
\end{axopicture}
\vspace{30pt} 
$ $ \hfill
\end{center}
\end{widetext}
\vspace{10pt}
plus a graph where the intermediate charged Higgs is replaced by a $W$ and removing the intermediate photon, graphs 
where the external charged Higgs and the $W$ are interchanged (also removing the photon), 
a graph where $(A,J)$ is replaced by a $Z$, graphs that involve quartic couplings, 
and a graph with a neutrino in the $t-$channel. This mode is dominated by the graph where the charged Higgs 
is on-shell. Note that the coupling $ZH^+W^-$ is suppressed by the triplet vev. This mode is 
enhanced when $H^-$ is also on-shell, corroborated by the fact that $B(H^-\rightarrow h_1W^-)=0.8$ in B2.

Similarly, in Fig.~\ref{Hp_prod} we see that the mode
$e^+e^- \rightarrow H^+ W^+ \Delta^{--}$ dominates in B3. It is represented by,
\begin{widetext}
\begin{center}
\vspace{-50pt} \hfill \\
\begin{axopicture}(380,100)(0,26) 
\ArrowLine(20,60)(50,30)
\ArrowLine(50,30)(20,0)
\Photon(50,30)(90,30){-3}{5}
\DashLine(120,60)(90,30){4}
\Photon(90,30)(120,0){-3}{5}
\Text(10,60)[]{$e^-$}
\Text(10,0)[]{$e^+$}
\Text(70,40)[]{$Z$}
\Text(100,50)[]{$H^-$}
\Text(135,2)[]{$W^+$}
\DashLine(120,60)(150,80){4}
\DashLine(150,40)(120,60){4}
\Text(160,80)[]{$H^+$}
\Text(165,40)[]{$\Delta^{--}$}
\Text(180,30)[]{$+$}
\ArrowLine(210,60)(240,30)
\ArrowLine(240,30)(210,0)
\Photon(240,30)(280,30){-3}{5}
\DashLine(310,60)(280,30){4}
\DashLine(280,30)(310,0){4}
\Text(200,60)[]{$e^-$}
\Text(200,0)[]{$e^+$}
\Text(260,40)[]{$Z,\gamma$}
\Text(289,50)[]{$\Delta^{++}$}
\Text(325,2)[]{$\Delta^{--}$}
\DashLine(310,60)(340,80){4}
\Photon(340,40)(310,60){-3}{5}
\Text(350,80)[]{$H^+$}
\Text(355,40)[]{$W^+$}
\end{axopicture}
\vspace{30pt} 
$ $ \hfill
\end{center}
\end{widetext}
\vspace{10pt}
plus a graph where the external particles $H^+$ and $\Delta^{--}$ are interchanged and 
at the same time the intermediate 
$\Delta^{++}$ is replaced by $H^-$, plus two graphs where the $H^-$ is replaced by a $W^-$ with $Z$ exchanged 
for a photon, and two graphs with quartic couplings. As it was mentioned before, the production
of a $\Delta^{++}$ is important because it could lead to the observation of its decay into two 
charged leptons, which could probe the mechanism for neutrino masses.

%
\begin{figure*}
\includegraphics[width=\textwidth]{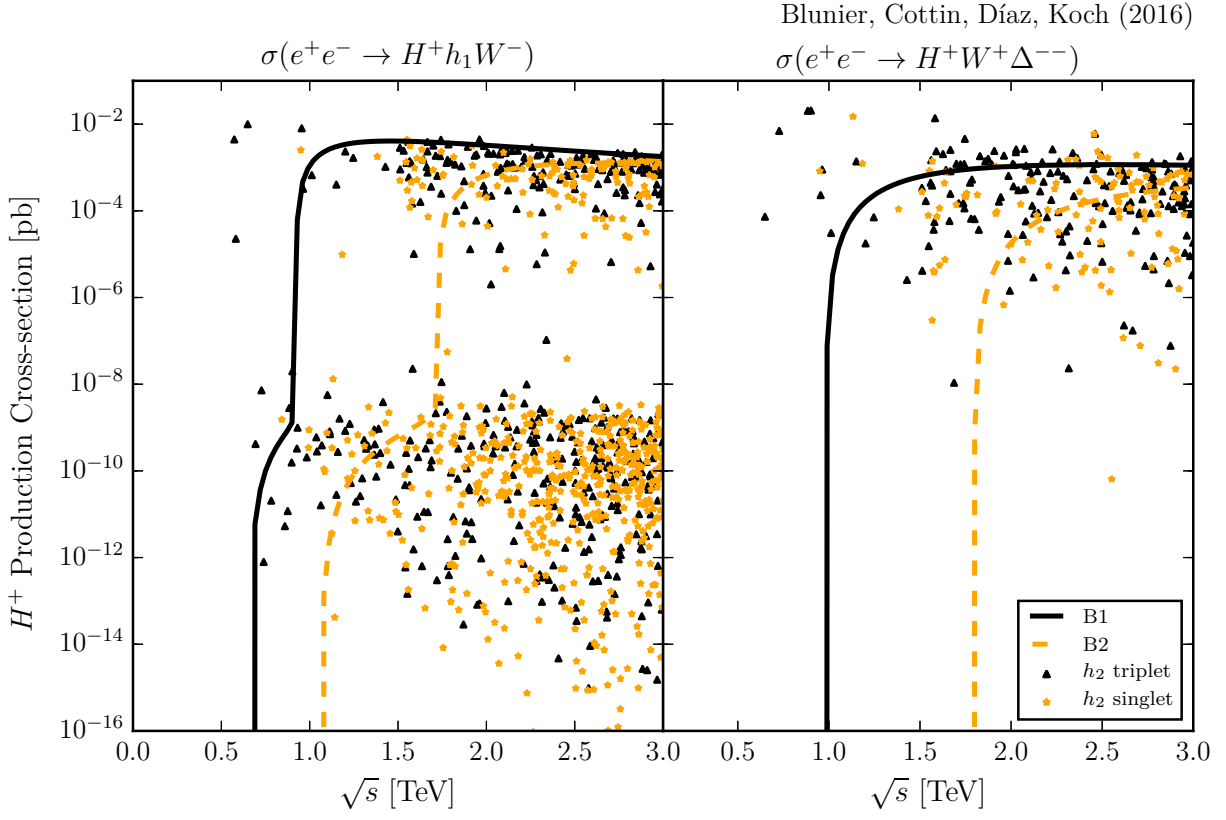}
\caption{Production modes $e^+e^-\rightarrow H^+h_1W^-$ and $e^+e^-\rightarrow H^+W^+\Delta^{--}$.}
\label{cloud_Hp_prod}
\end{figure*}
%

Fig.~\ref{cloud_Hp_prod} shows a general scan for the 3-body production modes
$e^+e^-\rightarrow H^+h_1W^-$ (left frame) and $e^+e^-\rightarrow H^+W^+\Delta^{--}$ (right frame).
For the case $e^+e^-\rightarrow H^+h_1W^-$, the mayority of the scenarios give a cross-section between 
$10^{-2}$ and $10^{-4}$ pb, as long as a second heavy particle is also on-shell. In the case 
of $e^+e^-\rightarrow H^+W^+\Delta^{--}$, the cross-section is of the same order 
between $10^{-3}$ and $10^{-5}$ pb, also independent of the composition of $h_2$. If neutrinos acquire their 
mass via a coupling to the triplet, the mechanism can be probed through the production of a double 
charged Higgs boson.

\section{Decay Branching Fractions}
\label{decays}

In this Section, we study the decay modes of the SM-like Higgs boson $h_{1}$, the next-to 
heaviest Higgs $h_{2}$, the $CP-$odd Higgs $A$, and the charged Higgs $H^{+}$. For the 
computation of branching fractions, we consider $B=\Gamma(H\rightarrow (XX)_{i})/\sum_{i} \Gamma(H\rightarrow (XX)_{i})$, with $H=h_{1}, h_{2}, A, H^{\pm}$. For the $CP-$even Higgses we have $XX=\tau\bar{\tau}, b\bar{b}, WW, ZZ, \gamma\gamma, Z\gamma, gg, JJ, JZ$ for $h_{1}$ and we include $t\bar{t}$ and $h_{1}h_{1}$ to the previous list for $h_{2}$. For $A$ we consider $XX=\tau\bar{\tau}, b\bar{b}, t\bar{t}, h_{i}Z, h_{i}J,\gamma\gamma,Z\gamma, gg$, with $i=1,2$. For $H^{\pm}$, we have $XX=t\bar{b}, h_{i}W^{\pm}, JW^{\pm}, ZW^{\pm}$, with $i=1,2$. 

We define
\begin{equation}
\lambda(a,b,c) = a^2+b^2+c^2-2ab-2ac-2bc .
\label{LambdaFunction}
\end{equation}
In the special case $b=c$, it is reduced to the function $\beta$,
\begin{equation}
\beta(b/a) = \frac{1}{a} \lambda^{1/2}(a,b,b) = \sqrt{1-4\frac{b}{a}}.
\label{BetaFunction}
\end{equation}
%

\subsection{$h_1$ and $h_2$ Decays}
\label{h1decays}

We first mention the decay modes to fermions for $h_{i}$ ($i=1,2$), which include 
$h_{i}\rightarrow b\bar{b}$ and $h_{i}\rightarrow \tau\bar{\tau}$. The decay $h_{2}\rightarrow t\bar{t}$ 
is considered for $h_{2}$, but not for $h_{1}$. The corresponding Feynman diagram is
\begin{center}
\vspace{-50pt} \hfill \\
\begin{axopicture}(90,90)(0,26) 
\DashLine(0,30)(40,30){4}
\ArrowLine(70,60)(40,30)
\ArrowLine(40,30)(70,0)
\Text(10,40)[]{$h_i$}
\Text(80,60)[]{$\bar{f}$}
\Text(80,2)[]{$f$}
\end{axopicture}
\vspace{30pt} 
$ $ \hfill
\end{center}
\vspace{10pt}
with Feynman rule given in Appendix \ref{FeynmanRules}.

The decay widths are given by
\begin{equation}
\Gamma(h_{i}\rightarrow f\bar{f})=\frac{N_c m_{h_i}}{8\pi}
\beta^3(m^2_{f}/m^2_{h_i}) |\lambda_{h_i ff}|^2,
\end{equation}
where the number of colors is $N_{c}=3$ for quarks and $N_{c}=1$ for leptons. We define the coupling 
$\lambda_{h_i ff}=O^{i2}_{\chi}h_f/\sqrt{2}$, where  $h_f$ corresponds to the respective Yukawa coupling 
in the convention $m_f=h_f v_\phi/\sqrt{2}$.

Since $h_1$ is always mainly doublet and $h_2$ is not, decay rates of $h_1$ to fermions are consistently 
larger than decay rates of $h_2$ to fermions. Similarly, since the $h_2$ component to doublet is larger in
B2 compared to B1 and B3, the corresponding decay rate is larger too.

Also important are the vector boson decays $h_{i}\rightarrow W^{+}W^{-}$, $h_{i}\rightarrow ZZ$,
with Feynman diagram,
\begin{center}
\vspace{-50pt} \hfill \\
\begin{axopicture}(90,90)(10,23) 
\DashLine(0,30)(40,30){5}
\Photon(70,60)(40,30){3}{5}
\Photon(70,0)(40,30){-3}{5}
\Text(5,40)[]{$h_i$}
\Text(90,60)[]{$Z,W$}
\Text(90,2)[]{$Z,W$}
\end{axopicture}
\vspace{30pt}
$ $ \hfill
\end{center}
\vspace{10pt}
The decay rate where both gauge bosons are on-shell is
\begin{widetext}
\begin{equation}
\Gamma(h_{i}\rightarrow VV) = \frac{m^3_{h_i} \delta'_{V}}{128\pi m^4_V}
\left[1-\frac{4m^{2}_{V}}{m^{2}_{h_{i}}}+\frac{12m^{4}_{V}}{m^{4}_{h_{i}}}\right]
\beta(m^2_V/m^2_{h_i}) |M_{h_i VV}|^2,
\label{HtoVVstar}
\end{equation}
\end{widetext}
with $V=Z,W$, $\delta'_{W}=2$ and $\delta'_{Z}=1$. The decay rate where one vector boson is off-shell is
\begin{equation}
\Gamma(h_{i}\rightarrow VV^{*}) = 
\frac{3 g^2_V m_{h_{i}}\delta_{V}}{512\pi^3m^2_{V}} F(m_V/m_{h_i})
|M_{h_{i}VV}|^2,
\label{HtoVV}
\end{equation}
with $g_{W}=g$, $g_{Z}=g/c_W$, $\delta_W=1$,
and $\delta_Z=\frac{7}{12}-\frac{10}{9}s^2_W+\frac{40}{27}s^4_W$, where $s_W$ and $c_W$ are the sine 
and cosine of the Weinberg angle. 
The $F$ function is defined 
in \cite{Gunion:1989we}. The relevant couplings (with units of mass) can be read 
from Appendix \ref{FeynmanRules}, from where we define
\begin{align}
M_{h_i WW}=&\,\frac{1}{2} g^2 (O^{i2}_{\chi}v_{\phi}+2 O^{i3}_{\chi}v_{\Delta}),
\label{hiWW}\\
M_{h_i ZZ}=&\,\frac{1}{2} (g^2+g'^2) (O^{i2}_{\chi}v_{\phi}+4O^{i3}_{\chi}v_{\Delta}),
\end{align}
and use them in eq.~(\ref{HtoVVstar}) and eq.~(\ref{HtoVV}).
In the case of $h_2$, since the penalization due to vev is already large ($v_\Delta/v_\phi\sim10^{-3}$
for our benchmarks), the $h_2$ component to doublet becomes important. Thus, the couplings 
$h_2VV$ are larger for B2, and in turn the decay rate (and branching fractions).

The decay to $\gamma\gamma$ is given by~\cite{Carena:2012xa,Arbabifar:2012bd},
\begin{align}
\Gamma(h_i \rightarrow \gamma\gamma)&=
\frac{\alpha^2g^2}{1024\pi^3} \frac{m^3_{h_i}}{m^2_W}
\Big| F_0(\tau_{H^+}^i) \frac{m_W}{m^2_{H_+}} M_{h_i H^+ H^-} \nonumber\\
&+4F_0(\tau_{\Delta}^i) \frac{m_W}{m^2_{\Delta^{++}}} M_{h_i \Delta^{++}\Delta^{--}}
\nonumber\\ 
&+ \, F_1(\tau_W^i) \frac{1}{m_W} M_{h_iWW} \nonumber\\
&+\frac{4\sqrt{2}}{3h_t} F_{1/2}(\tau_t^i) \lambda_{h_i t t} \Big|^2,
\label{GammaPhotonPhoton}
\end{align}
where the couplings $M_{h_i H^+ H^-}$ (in our convention $H^+\equiv h_2^+$), 
$M_{h_i \Delta^{++}\Delta^{--}}$, and $M_{h_iWW}$ are defined in Appendix \ref{FeynmanRules}
and in eq.~(\ref{hiWW}).
In eq.~(\ref{GammaPhotonPhoton}) we have defined
$\tau_a^i=4m^2_a/m^2_{h_i}$ where $a=H^+,\Delta,W$. The $F_0, F_1$ and $F_{1/2}$ functions are defined 
in \cite{Gunion:1989we}. 

The decay to $Z\gamma$ is given by~\cite{Carena:2012xa,Arbabifar:2012bd}
\begin{equation}
\Gamma(h_{i}\rightarrow Z \gamma) = 
\frac{\alpha g^{2}}{2048\pi^{4}m^{4}_{W}}|A|^2m^3_{h_{i}}(1-\frac{m^2_{Z}}{m^2_{h_{i}}})^3,
\end{equation}
where $A$ is defined as
\begin{equation}
A= A_{W}+A_{t}+A^{H+}_{0}+2A^{\Delta^{++}}_{0},
\end{equation}
with
\begin{widetext}
\begin{eqnarray}
A_W+A_t &=&
c_W \, M_{h_i WW} \, A_1(\tau_W,\lambda_W) +
\frac{g m_W}{c_W} N_cQ_t (1-4Q_ts^2_W) \, \lambda_{h_i tt} \, A_{1/2}(\tau_t, \lambda_t)
\nonumber\\
A^{H^+}_0 &=&
\frac{m^2_W}{gs_Wm^2_{H^+}} \, \lambda_{ZH^+H^-} \, M_{h_iH^+H^-} \,
A_0(\tau_{H^+}, \lambda_{H^+})
\nonumber\\
A^{\Delta^{++}}_0 &=&
\frac{m^2_W}{gs_Wm^2_{\Delta^{++}}} \,
\lambda_{Z\Delta^{++}\Delta^{--}} \, M_{h_i\Delta^{++}\Delta^{--}} \,
A_0(\tau_{\Delta^{++}}, \lambda_{\Delta^{++}}),
\label{eq:At}
\end{eqnarray}
\end{widetext}
where
\begin{eqnarray}
\lambda_{ZH^+H^-} &=& -\frac{g}{2c_W}(s_\beta^2-2s_W^2),
\nonumber\\
\lambda_{Z\Delta^{++}\Delta^{--}} &=& -\frac{g}{c_W}(c_W^2-s_W^2), 
\end{eqnarray}
as can be seen from Appendix \ref{FeynmanRules}. The loop functions are,
\begin{widetext}
\begin{eqnarray}
A_0(\tau,\lambda) &=& I_1(\tau,\lambda),
\nonumber\\
A_1(\tau,\lambda) &=& 4(3-\tan^2{\theta_W})I_{2}(\tau,\lambda) +
[(1+2/\tau)\tan^2{\theta_W}-(5+2/\tau)]I_{1}(\tau,\lambda),
\nonumber\\
A_{1/2}(\tau,\lambda) &=& I_1(\tau,\lambda)-I_2(\tau,\lambda),
\end{eqnarray}
\end{widetext}
with $\tau_{b}=\frac{4m^2_{b}}{m^2_{h_{i}}}$, $\lambda_{b}=\frac{4m^2_{b}}{m^2_{Z}}$, 
$b=t,W,H^{+},\Delta^{++}$, and the parametric integrals $I_{1}, I_{2}$ are specified in 
\cite{Gunion:1989we}. 

We also consider the 1-loop decay to $gg$ for completeness. It is given by~\cite{Gunion:1989we}
\begin{equation}
\Gamma(h_{i}\rightarrow gg) = 
\frac{\alpha^2_{s} g^{2}m^{3}_{h_{i}}}{128\pi^3m^2_{W}} \Big|\frac{4\sqrt{2}}{3h_t} F_{1/2}(\tau_t^i) \lambda_{h_i t t} \Big|^2
\end{equation}
with the $F_{1/2}$ given in Appendix C of~\cite{Gunion:1989we}.

The decay to Majorons $h_i\rightarrow JJ$ and $h_i\rightarrow J Z$ proceeds
with a negligible Majoron mass. The decay rates are given by,
\begin{equation}
\Gamma(h_i\rightarrow JZ) =
\frac{m^3_{h_i}}{16 \pi m^2_Z} |\lambda_{Zh_iJ}|^2 \left( 1-\frac{m^2_Z}{m_{h_i}^2} \right)^3 
\end{equation}
and
\begin{equation}
\Gamma(h_i\rightarrow JJ) = \frac{|M_{h_iJJ}|^2}{32\pi m_{h_i}},
\end{equation}
with
\begin{equation}
\lambda_{Zh_iJ} = \frac{g}{2c_W} (O_\chi^{i2}O_\varphi^{22}-2O_\chi^{i3}O_\varphi^{23}).
\end{equation}
$M_{h_iJJ}$ is defined from the corresponding Feynman rule in Appendix~\ref{FeynmanRules}.

Finally, the decay $h_2\rightarrow h_1 h_1$ is given by,
\begin{equation}
\Gamma(h_2\rightarrow h_1h_1) = \frac{\beta(m^2_{h_{1}}/m^2_{h_{2}})} {32 \pi m_{h_{2}}} |M_{h_2h_1h_1}|^2,
\end{equation}
where $M_{h_2h_1h_1}$ is defined from the corresponding Feynman rule in Appendix~\ref{FeynmanRules}.

In the case of $h_1$ we require that its mass is $\approx125$ GeV and that it is mostly doublet. 
Besides the usual decay modes for this SM-like Higgs boson, in this model there are two more. These are
$h_{1}\rightarrow JJ$ and $h_{1}\rightarrow JZ$. For the three benchmarks, the branching fractions are
B$(h_{1}\rightarrow JJ)\approx3\times10^{-5}$ and B$(h_{1}\rightarrow JZ)\approx3\times10^{-13}$. We are 
well within experimental constraints on the Higgs invisible width, as branching fractions bigger than $22\%$
are excluded at 95\% CL~\cite{Falkowski:2013dza}. These modes 
are suppressed due to two different reasons. The mode $h_1\rightarrow JZ$ is 
suppressed because the Majoron $J$ is mostly singlet. The decay mode $h_1\rightarrow JJ$ is 
suppressed because in addition we require a small value for $\beta_2$. 
%
\begin{figure*}
\includegraphics[width=\textwidth]{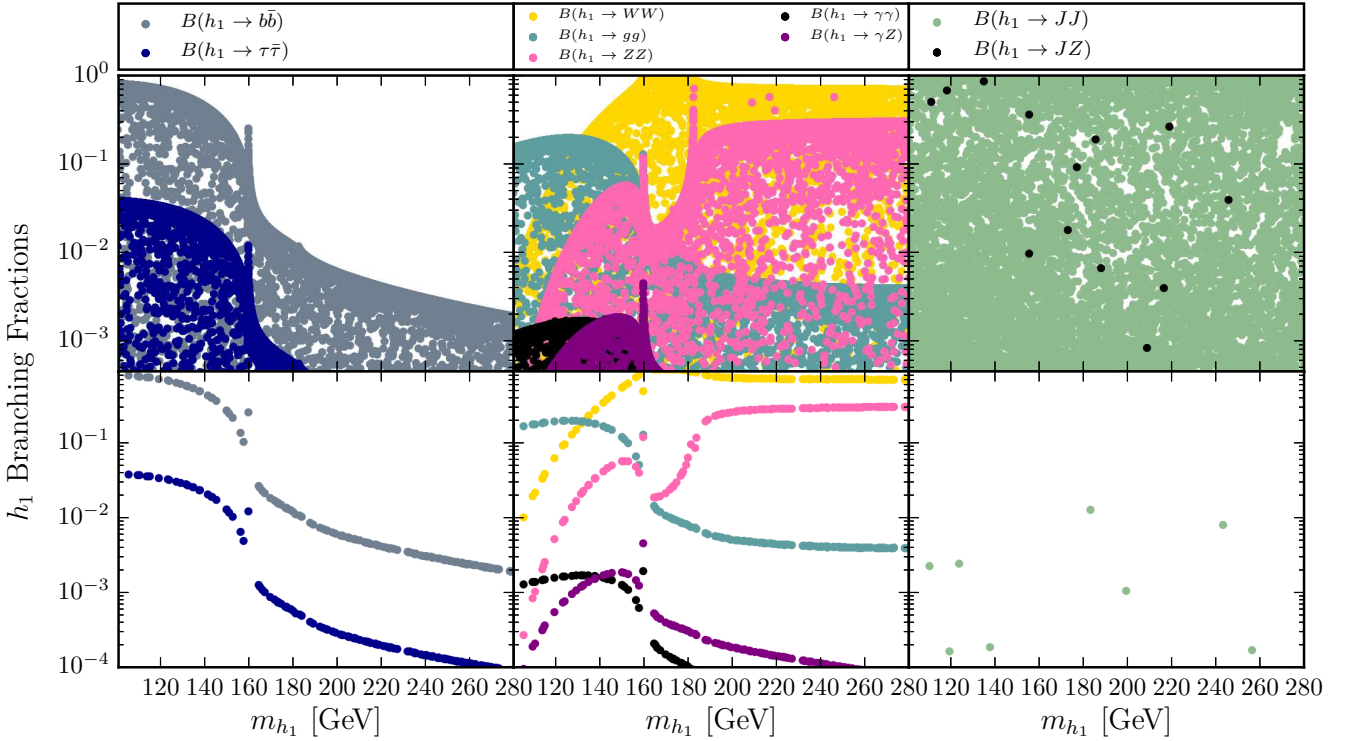}
\caption{ Branching fractions for the $h_1$ scalar with (bottom) and without (top) restrictions, as explained in the text.}
\label{BRh1Vsmh1_scan}
\end{figure*}
%

Fig.~\ref{BRh1Vsmh1_scan} shows the branching fractions of our light Higgs $h_1$. In the top
frame we scan the parameters without any restriction, varying $\lambda_1$ between $[0,4]$, 
in order not to constrain the Higgs mass, as we need to make sure the points in the plot are 
consistent with a SM-like Higgs. Also is useful to keep the mass free to observe the effect 
of the constraints and to facilitate the comparison with $h_2$. 
On the top frame $\beta_2$ is not constrained and varies between $[\text{-}4,4]$ so we can clearly see
the suppression in the Majoron decays once we constrain its value in the bottom frame. The
bottom frame includes all constrains from Section \ref{ParamRestrictions}.
\begin{table}
\caption{Branching fractions for $h_2$ in the three different benchmarks.
\label{tab5}}
\begin{tabular*}{0.5\textwidth}{@{\extracolsep{\fill}} c c c c }
\hline\hline
Branching Fraction               & B1                  & B2                     & B3        \\ \hline
B$(h_2\rightarrow t\bar{t})$     & $0.3$ 			   & $7.9\times10^{-3}$    & -                   \\
B$(h_2\rightarrow b\bar{b})$     & $6.0\times10^{-4}$  & $9.5\times10^{-6} $   & $3.4\times10^{-7}$ \\
B$(h_2\rightarrow \tau\tau)$     & $3.0\times10^{-5}$  & $4.5\times10^{-7}$    & $1.6\times10^{-8}$ \\
B$(h_2\rightarrow WW)$           & $7.0\times10^{-3}$  & $3.0\times10^{-2}$    & $3.6\times10^{-6} $ \\
B$(h_2\rightarrow ZZ)$           & $0.6$               & $1.0\times10^{-2}$    & $1.3\times10^{-4} $ \\
B$(h_2\rightarrow gg)$           & $7.2\times10^{-3}$  & $1.3\times10^{-4}$    & $1.0\times10^{-6} $ \\
B$(h_2\rightarrow \gamma\gamma)$ & $7.7\times10^{-6}$  & $2.9\times10^{-5}$    & $1.8\times10^{-3} $ \\
B$(h_2\rightarrow Z\gamma)$      & $1.6\times10^{-6}$  & $1.6\times10^{-7}$    & $1.9\times10^{-7} $ \\
B$(h_2\rightarrow JJ)$           & $1.2\times10^{-4}$  & $0.9$                 & $0.9$               \\
B$(h_2\rightarrow JZ)$           & $3.0\times10^{-2}$  & $3.6\times10^{-12} $  & $2.5\times10^{-6} $ \\
B$(h_2\rightarrow h_1h_1)$       & $0.1$               & $1.7\times10^{-2}$    & $1.0\times10^{-6} $ \\
\hline\hline
\end{tabular*}
\end{table}
The branching fractions in our three benchmarks for $h_2$ are given 
in Table \ref{tab5}. We mention first that $h_2$ has a larger doublet 
component in B2, and for that reason decay rates to fermions are larger 
in that benchmark. Nevertheless, this fact is obscured in branching 
fractions because the total decay rate is also very different. Similarly, decay 
rates to gauge bosons are larger in B2, but not necessarily the same is true at the level 
of branching fractions. Clearly, looking at branching fractions, decays of $h_2$ to two Majorons 
(invisible decay) dominate in B2 and B3 because $h_2$ has a large singlet component in 
those two benchmarks.

\begin{figure*}
\includegraphics[width=\textwidth]{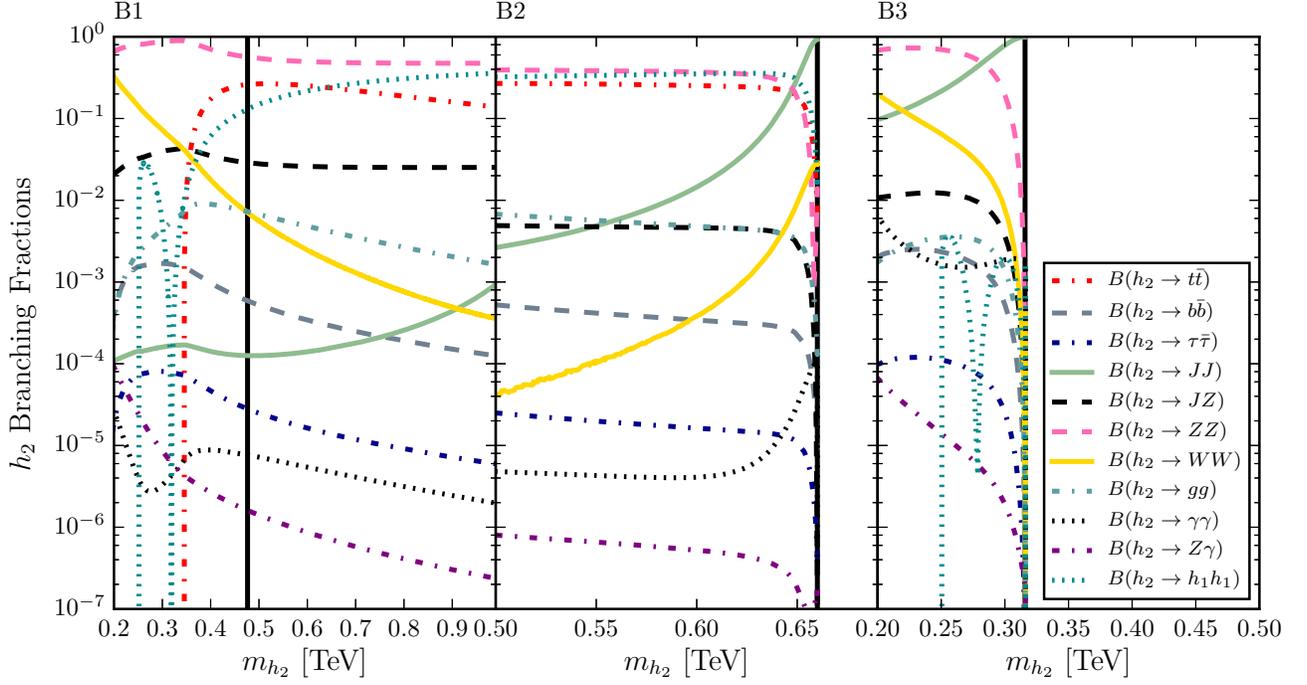}
\caption{Branching fractions for the $h_2$ scalar in the three benchmarks as a 
function of $m_{h_{2}}$. The parameter $\kappa$ is varied to 
move $m_{h_{2}}$, as explained in the text. The vertical solid line 
in each frame corresponds to our benchmark point. The plot includes all constrains 
from Section \ref{ParamRestrictions}.}
\label{BRAVsh2_Curves}
\end{figure*}

\begin{figure*}
\includegraphics[width=\textwidth]{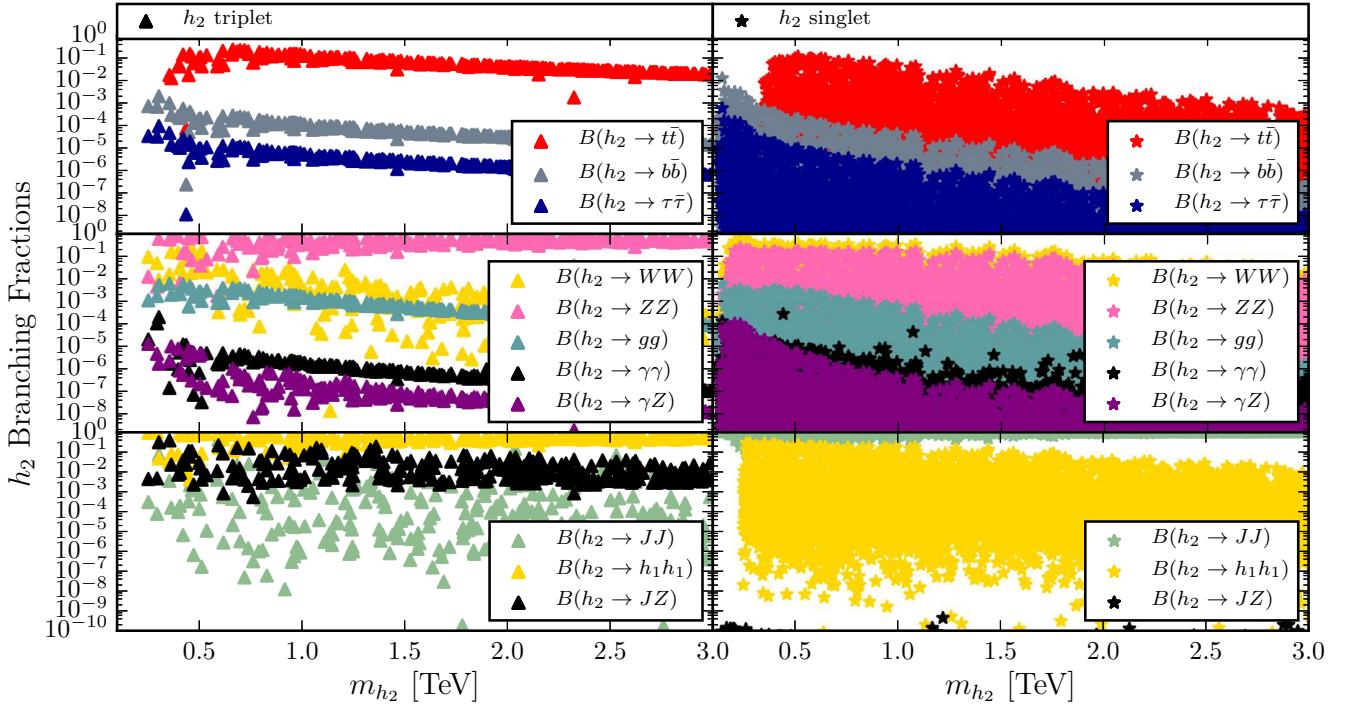}
\caption{Branching fractions for the $h_2$ scalar as a function of $m_{h_{2}}$. The left column shows 
points where $h_{2}$ is triplet-like (i.e $|O^{23}_{\chi}|>0.95$). The right column shows 
points where $h_{2}$ is singlet-like (i.e $|O^{21}_{\chi}|>0.95$). Parameters are 
varied according to Table \ref{tab1}. The scan includes all constrains from Section \ref{ParamRestrictions}. 
}
\label{BRh2Vsmh2_scan}
\end{figure*}

Fig.~\ref{BRAVsh2_Curves} shows the branching fractions as a function of the scalar mass 
$m_{h_2}$, evolving from our three benchmarks, while Fig.~\ref{BRh2Vsmh2_scan} shows a scan of the $h_2$ 
decays, with all the constrains from Section \ref{ParamRestrictions} implemented.

%
%
The curves shown in Fig.~\ref{BRAVsh2_Curves} confirms the previous observations. These
curves are found by keeping the values of the independent parameters as in the 3 different benchmarks,
and varying the value of $\kappa$ in order to keep $m_{h_2}$ free. Since due to mixing this procedure 
will also vary the value of $m_{h_1}\approx125$ GeV, we keep $\lambda_1$ also free to 
compensate, as in Table \ref{tab1}. We show also as a vertical solid line the 
value of $m_{h_2}$ in the corresponding benchmark. In the case
of B2, near the vertical line $h_2$ is mainly singlet, and $\kappa$ affects very little to 
$m_{h_2}$. If $\kappa$ is suficiently different from its starting value in B2
$h_2$ becomes mostly triplet. The value for $m_{h_2}$
cannot be larger than its value in the benchmark because by then $h_2$ is mostly singlet and $\kappa$
has little effect. Something similar happens with B3. In all cases $h_2\rightarrow ZZ$ and
$h_2\rightarrow WW$ are important. 
Decays to fermions depend strongly on the (small) $h_2$ component to doublet.
%
%
In the scan in Fig.~\ref{BRh2Vsmh2_scan}, we plot $h_{2}$ branching fractions 
while all the parameters are varied according to Table \ref{tab1}. We see that the values of 
the branching fractions separates in two regions, that we plot separately in the two 
column plot. These two sectors corresponds to a mainly triplet (left column) or mainly 
singlet (right column) $h_2$. The scan shows that if $h_2$ is mainly 
triplet (as in B1) decay modes $h_2\rightarrow ZZ$ and $h_2\rightarrow h_1 h_1$
can dominate, with $h_2\rightarrow JZ$ sometimes also important. On the
contrary, if $h_2$ is mainly singlet (as in B2) the decay mode $h_2\rightarrow JJ$ 
dominates by far, with $h_2\rightarrow WW$ and $h_2\rightarrow ZZ$ following in importance.
The $h_2\rightarrow t\bar{t}$ branching fractions can be large as long as the other decay rates are also small.

\subsection{$A$ Decays}

Now we study the decays of the $CP-$odd Higgs boson $A$. The relevant decays at tree-level 
are to third generation fermions, $A\rightarrow t\bar{t}, A\rightarrow b\bar{b} , A\rightarrow \tau\tau$, 
to $CP-$even Higgs bosons and a Majoron, $A\rightarrow h_{i} J$, and to $CP-$even Higgs bosons and a $Z$ gauge
boson, $A\rightarrow h_i Z$. We also consider the 1-loop decays to $\gamma\gamma$, $Z\gamma$ and $gg$ for completeness.

The decay of $A$ to fermions, represented by the Feynman diagram,
\begin{center}
\vspace{-50pt} \hfill \\
\begin{axopicture}(90,90)(0,26) 
\DashLine(0,30)(40,30){4}
\ArrowLine(70,60)(40,30)
\ArrowLine(40,30)(70,0)
\Text(20,40)[]{$A$}
\Text(80,60)[]{$\bar{f}$}
\Text(80,2)[]{$f$}
\end{axopicture}
\vspace{30pt} 
$ $ \hfill
\end{center}
\vspace{10pt}
is given by
\begin{equation}
\Gamma(A\rightarrow f\bar{f}) = \frac{N_c m_A}{8\pi}\left[1-4 \frac{m^2_f}{m_A^2}\right]^{\frac{1}{2}}|\lambda_{Aff}|^2,
\end{equation}
with a coupling
\begin{equation}
\lambda_{Aff} = \frac{1}{\sqrt{2}} O^{32}_{\varphi} h_{f},
\end{equation}
as seen in Appendix~\ref{FeynmanRules}. $h_{f}$ is the Yukawa coupling of the fermion. 
Since $A$ is always mainly triplet, $O^{32}_{\varphi}$ is always small. The decay $A\rightarrow f\bar{f}$
proceeds just because the $A$ eigenfunction has a small component of doublet, as indicated in eq.~(\ref{NeuFields123Model}).

The $A$ boson can also decay into a $CP-$even Higgs and a $Z$ boson. The corresponding Feynman diagram is,
\begin{center}
\vspace{-50pt} \hfill \\
\begin{axopicture}(90,90)(0,26) 
\DashLine(0,30)(40,30){4}
\DashLine(40,30)(70,60){4}
\Photon(40,30)(70,0){-3}{5}
\Text(20,40)[]{$A$}
\Text(80,60)[]{$h_i$}
\Text(80,2)[]{$Z$}
\end{axopicture}
\vspace{30pt} 
$ $ \hfill
\end{center}
\vspace{10pt}
The decay rate is given by the formula,
\begin{equation}
\Gamma(A\rightarrow h_{i}Z) = 
\frac{\lambda_{Ah_iZ}^2}{16\pi} \, \frac{m_A^3}{m^2_Z} \,
\lambda^{3/2}\left(1,m_{h_i}^2/m_A^2,m_Z^2/m_A^2\right),
\end{equation}
with a coupling
\begin{equation}
\lambda_{Ah_iZ} = \frac{g}{2c_W}(O_\chi^{i2}O_\varphi^{32}-2O_\chi^{i3}O_\varphi^{33}),
\end{equation}
as seen in Appendix \ref{FeynmanRules}. The $\lambda$ function is defined in eq.~(\ref{LambdaFunction}).
In the case $A\rightarrow h_2Z$, since $A$ is always mainly
triplet, there is no phase space in B1, where $h_2$ is also a triplet and has a mass almost equal 
to the mass of $A$. In the case $A\rightarrow h_1Z$, since the couplings are more or less similar for B1 and B2, the difference
is due to the value of $m_A$.

The decay to a $CP-$even Higgs boson and a Majoron is represented by the following Feynman diagram,
\begin{center}
\vspace{-50pt} \hfill \\
\begin{picture}(90,90)(0,26) 
\DashLine(0,30)(40,30){4}
\DashLine(40,30)(70,60){4}
\DashLine(40,30)(70,0){4}
\Text(10,40)[]{$A$}
\Text(80,60)[]{$h_i$}
\Text(80,2)[]{$J$}
\end{picture}
\vspace{30pt} 
$ $ \hfill
\end{center}
\vspace{10pt}
The decay rate is
\begin{equation}
\Gamma(A\rightarrow h_i J) = \frac{M^2_{h_i a_1 a_2}}{16 \pi m_A}
\lambda^{1/2}(1,m_{h_i}^2/m_A^2,m_J^2/m_A^2),
\end{equation}
with the coupling $M_{h_i a_1 a_2}$ (with units of mass) given in Appendix \ref{FeynmanRules}.

The decay to $\gamma\gamma$ is given by~\cite{Gunion:1989we}

\begin{equation}
\Gamma(A\rightarrow \gamma\gamma) = \frac{\alpha^2 g^2 m^2_{A}}{1024 \pi^3 m^2_{W}}\Big|\frac{4\sqrt{2}}{3h_t} F_{1/2}(\tau_t) \lambda_{A t t} \Big|^2
\end{equation}

with $\tau_t=4m^2_{t}/m^2_{A}$ and the $F_{1/2}$ function for a pseudoscalar is defined in Appendix C of Ref.~\cite{Gunion:1989we}. 

The decay to $Z\gamma$ is given by~\cite{Gunion:1989we}

\begin{equation}
\Gamma(A\rightarrow Z \gamma) = 
\frac{\alpha g^{2}}{2048\pi^{4}m^{4}_{W}}|A_{t}|^2m^3_{A}(1-\frac{m^2_{Z}}{m^2_{A}})^3,
\end{equation}
where $A_{t}$ is defined in equation~\ref{eq:At} (replacing $h$ with $A$).

Finally, the decay to two gluons is~\cite{Gunion:1989we}

\begin{equation}
\Gamma(A\rightarrow gg) = 
\frac{\alpha^2_{s} g^{2}m^{3}_{A}}{128\pi^3m^2_{W}} \Big|\frac{4\sqrt{2}}{3h_t} F_{1/2}(\tau_t) \lambda_{A t t} \Big|^2.
\end{equation}

Branching fractions for the decay of $A$ for our three benchmarks are given in Table \ref{tab6}.
\begin{table}[h]
\caption{Branching fractions for $A$ in our three different benchmarks.
\label{tab6}}
\begin{center}
\begin{tabular*}{0.5\textwidth}{@{\extracolsep{\fill}} c c c c }
\hline\hline
Branching Fraction            & B1                  & B2                  & B3         \\ \hline
B$(A\rightarrow t\bar{t})$ & $0.5$              & $0.2$              &  -                 \\
B$(A\rightarrow b\bar{b})$ & $5.5\times10^{-4}$ & $1.5\times10^{-4}$ & $6.0\times10^{-3}$ \\
B$(A\rightarrow \tau\tau)$ & $2.6\times10^{-5}$ & $7.0\times10^{-6}$ & $2.8\times10^{-4}$ \\
B$(A\rightarrow h_1Z)$     & $0.5$              & $0.8$              & $0.9$              \\
B$(A\rightarrow h_1J)$     & $1.7\times10^{-2}$ & $4.4\times10^{-3}$ & $2.0\times10^{-2}$ \\
B$(A\rightarrow h_2Z)$     & -                  & $5.0\times10^{-2}$ & -                  \\
B$(A\rightarrow h_2J)$     & -                  & $1.1\times10^{-4}$ & -                  \\
B$(A\rightarrow gg)$            & $1.4\times10^{-2}$  & $2.7\times10^{-3}$ & $6.2\times10^{-2}$ \\
B$(A\rightarrow \gamma\gamma)$  & $1.7\times10^{-5}$  & $3.4\times10^{-6}$ & $7.7\times10^{-5}$  \\
B$(A\rightarrow Z\gamma)$       & $8.2\times10^{-7}$  & $2.6\times10^{-7}$ & $2.0\times10^{-6}$ \\
\hline\hline
\end{tabular*}
\end{center}
\end{table}
The $A$ boson component to doublet is the same for B1 and B2, but $m_A$ is not. This leads to 
larger decay rates to fermions in B2. Since the total decay rate is also different, this is not 
observed for branching fractions and in fact, the opposite happens. 
Note that in B1 and B3 the decays of $A$ to $h_{2}$ and a $J$ or a $Z$ 
are not kinematically allowed. The same happens in B3 for the decay to top quarks. 
In B2, $A$ can be much heavier than $h_2$ thus, the decay $A\rightarrow h_2Z$ is open.
%
\begin{figure*}
\includegraphics[width=\textwidth]{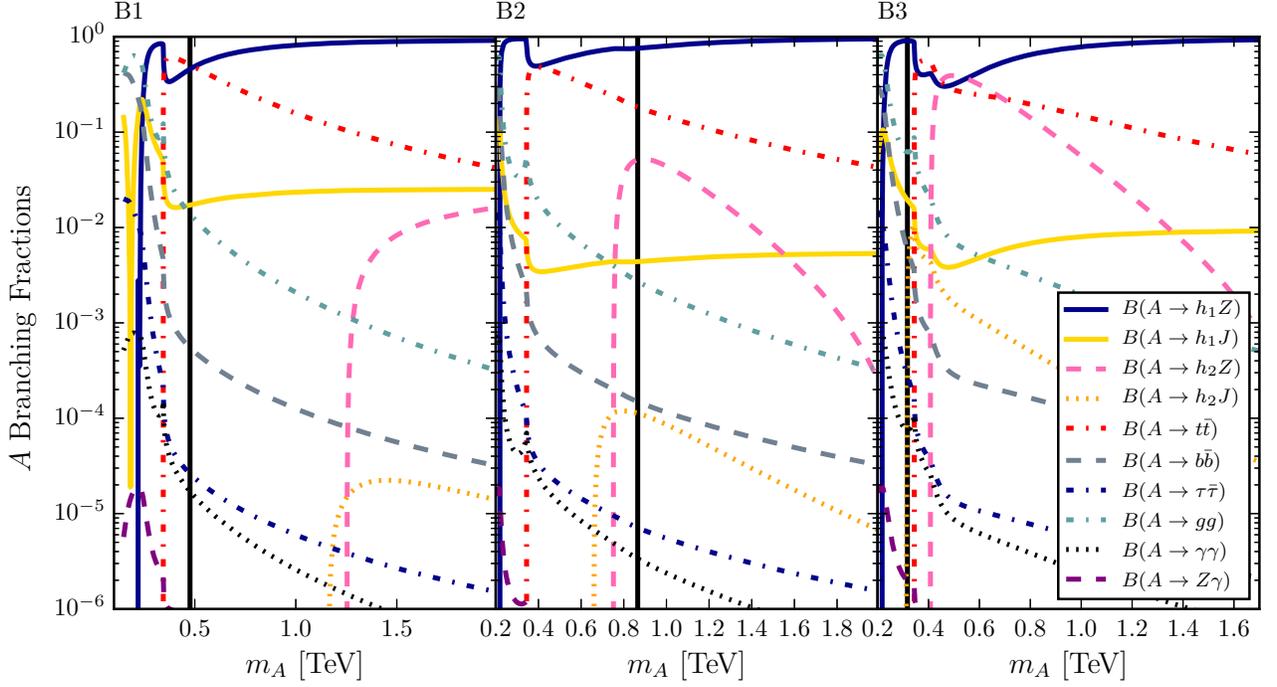}
\caption{$CP-$odd Higgs $A$ branching fractions in the three benchmarks as a 
function of $m_{A}$. The parameter $\kappa$ is varied to 
move $m_{A}$, as explained in the text. The vertical solid line 
in each frame corresponds to our benchmark point. The plot includes all constrains 
from Section \ref{ParamRestrictions}.
}
\label{BRAVsmA_Curves}
\end{figure*}
%

Fig.~\ref{BRAVsmA_Curves} shows the branching fractions of $A$ as a function of its mass. 
The curves are obtained starting from each of the 3 benchmarks and vary $\kappa$ to change $m_A$. Since
this procedure will also change $m_{h_1}$, which we want fixed to 125 GeV, we change also the 
value of $\lambda_1$ to recover $m_{h_1}\approx125$ GeV, as in Table \ref{tab1}. In all cases, the modes 
$A\rightarrow h_1Z$ and $A\rightarrow t\bar{t}$ dominate. In B3 the decay mode $A\rightarrow h_2Z$ is open and
can be relevant too. 
%
\begin{figure*}
\includegraphics[width=\textwidth]{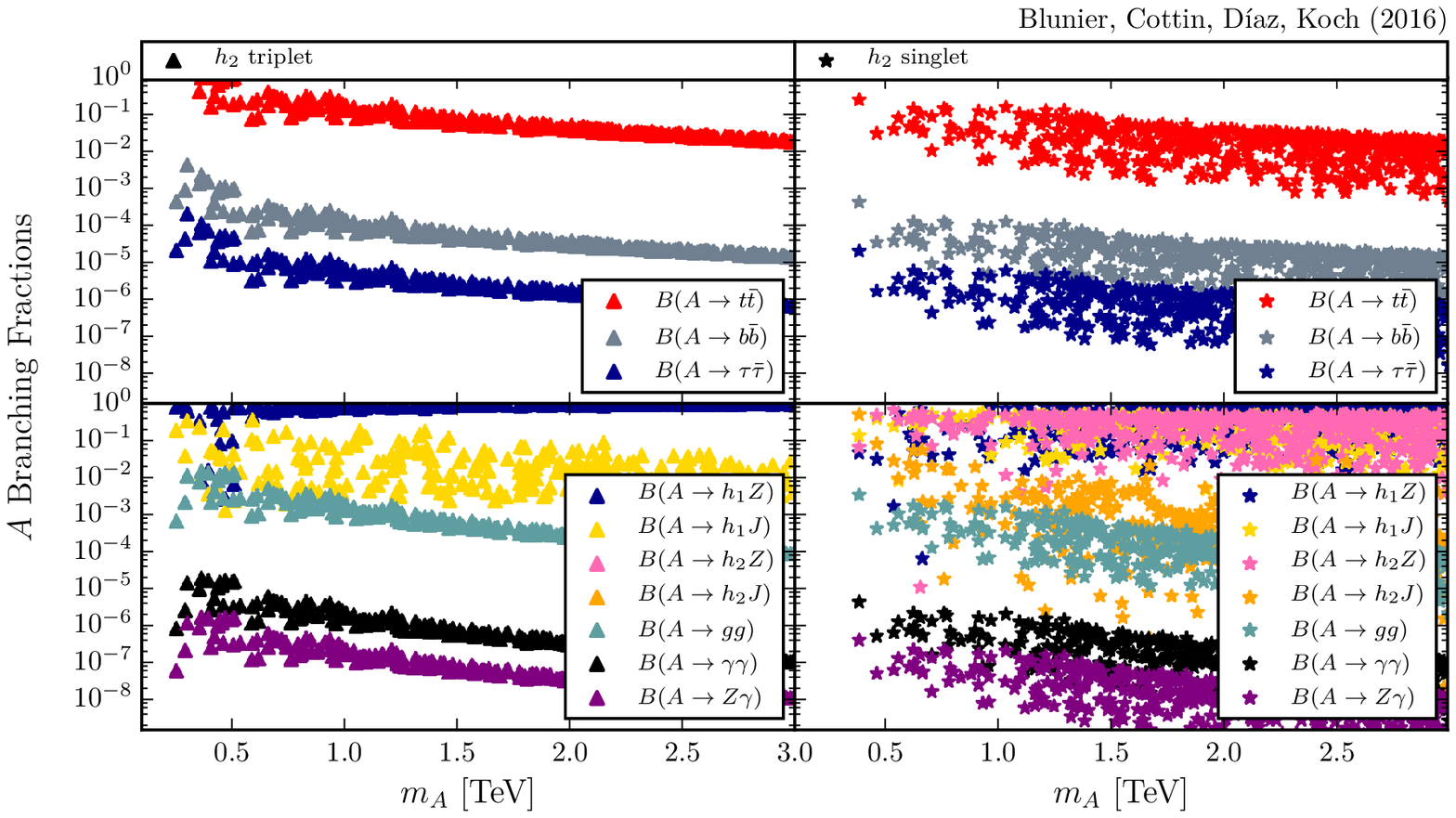}
\caption{Branching fractions for the $A$ scalar as a function of $m_{A}$. 
The left column shows points where $h_{2}$ is triplet-like (i.e $|O^{23}_{\chi}|>0.95$). The 
right column shows points where $h_{2}$ is singlet-like (i.e $|O^{21}_{\chi}|>0.95$). Parameters are 
varied according to Table~\ref{tab1}. The scan includes all constrains from 
Section~\ref{ParamRestrictions}.
}
\label{BRAVsmA_scan}
\end{figure*}
%

Fig.~\ref{BRAVsmA_scan} shows a general scan where all the parameters are 
varied according to Table~\ref{tab1}. It shows that the decay mode $A\rightarrow h_1Z$ 
dominates. If the channel is open, when $h_2$ is mainly singlet, 
the decay channel $A\rightarrow h_2Z$ is also very important.

\subsection{$H^\pm$ Decays}

In this Section we study tree-level decays of the singly charged Higgs boson.
The decay to $t\bar{b}$, represented by the Feynman diagram,
\begin{center}
\vspace{-50pt} \hfill \\
\begin{picture}(90,90)(0,26) 
\DashLine(0,30)(40,30){4}
\ArrowLine(70,60)(40,30)
\ArrowLine(40,30)(70,0)
\Text(20,40)[]{$H^+$}
\Text(80,60)[]{$\bar{b}$}
\Text(80,2)[]{$t$}
\end{picture}
\vspace{30pt} 
$ $ \hfill
\end{center}
%
has a rate
\begin{align}
\Gamma(H^{\pm}\rightarrow t\bar{b}) &= \frac{N_c (O_+^{21})^2}{16\pi m^3_{H^\pm}}
\Big[ (h^2_t+h^2_b) (m^2_{H^\pm}-m^2_t-m^2_b) \nonumber\\
&- 4 h_t h_b m_t m_b \Big] \lambda^{1/2}(m^2_{H^\pm},m^2_t,m^2_b).
\end{align}
Similarly, the decay $H^{\pm}\rightarrow h_iW^\pm$ 
%
%
\begin{center}
\vspace{-50pt} \hfill \\
\begin{axopicture}(90,90)(0,26) 
\DashLine(0,30)(40,30){4}
\DashLine(40,30)(70,60){4}
\Photon(40,30)(70,0){-3}{5}
\Text(10,40)[]{$H^+$}
\Text(80,60)[]{$h_i$}
\Text(82,2)[]{$W^+$}
\end{axopicture}
\vspace{30pt} 
$ $ \hfill
\end{center}
%
has a rate given by

\begin{align}
\Gamma(H^{\pm}&\rightarrow h_iW^\pm) = \frac{g^2 |\lambda_{H^\pm h_i W^\mp}|^2}{64\pi m^3_{H^+}m^2_W}
\lambda^{3/2}(m^2_{H^\pm},m^2_{h_i},m^2_W),
\end{align}

with,

\begin{equation}
\lambda_{H^{\pm} h_i W^{\mp}} = O^{21}_+ O^{i2}_\chi - \sqrt{2} O^{22}_+ O^{i3}_\chi.
\end{equation}
The decay to a Majoron and a $W^\pm$ boson is

\begin{center}
\vspace{-50pt} \hfill \\
\begin{picture}(90,90)(0,26) 
\DashLine(0,30)(40,30){4}
\DashLine(40,30)(70,60){4}
\Photon(40,30)(70,0){-3}{5}
\Text(10,40)[]{$H^+$}
\Text(75,60)[]{$J$}
\Text(82,2)[]{$W^+$}
\end{picture}
\vspace{30pt} 
$ $ \hfill
\end{center}
%
with a decay rate,
\begin{equation}
\Gamma(H^\pm\rightarrow JW^\pm) =
\frac{g^2 |\lambda_{H^\pm JW^\mp}|^2}{64\pi m^3_{H^+}m^2_W} [m^2_{H^\pm} - m^2_W]^3,
\end{equation}
where
\begin{equation}
\lambda_{H^\pm JW^\mp} = O^{21}_+ O^{22}_\varphi + \sqrt{2} O^{22}_+ O^{23}_\varphi,
\end{equation}
To finish, the decay to a $Z$ and a $W^{\pm}$ boson is,
\begin{center}
\vspace{-50pt} \hfill \\
\begin{axopicture}(90,90)(10,23) 
\DashLine(0,30)(40,30){5}
\Photon(70,60)(40,30){3}{5}
\Photon(70,0)(40,30){-3}{5}
\Text(10,40)[]{$H^+$}
\Text(80,60)[]{$Z$}
\Text(85,2)[]{$W^+$}
\end{axopicture}
\vspace{30pt}
$ $ \hfill
\end{center}
\vspace{30pt}
and has the following decay rate
\begin{widetext}
\begin{equation}
\Gamma(H^\pm\rightarrow ZW^\pm)= \frac{g^4 |M_{H^\pm Z W^\mp}|^2}
{256 \pi m^4_W m^3_{H^\pm} }\Big[ m^4_{H^\pm} + m^4_Z + 10m^2_Zm^2_W + m^4_W - 2m^2_{H^\pm}(m^2_W+m^2_Z) \Big]\lambda^{1/2}(m^2_{H^\pm}, m^2_Z, m^2_W),
\end{equation}
\end{widetext}
with
\begin{equation}
M_{H^\pm Z W^\mp} = O^{21}_+ s_W v_\phi - \sqrt{2} O^{22}_+ ( 1 + s^2_W) v_\Delta.
\end{equation}
In Table~\ref{tab7} we show the singly charged Higgs branching fractions in our three benchmarks.
\begin{table}[h]
\caption{Branching fractions for $H^{\pm}$ in our three benchmarks.
\label{tab7}}
\begin{tabular*}{0.5\textwidth}{@{\extracolsep{\fill}} c c c c }
\hline\hline
Branching Fraction                & B1       	    & B2                 & B3     \\ \hline
B$(H^\pm\rightarrow t\bar{b})$ & $7.0\times10^{-2}$ & $2.0\times10^{-2}$ & $0.2$ \\
B$(H^\pm\rightarrow h_1W^\pm)$ & $0.7$              & $0.8$              & $0.6$ \\
B$(H^\pm\rightarrow h_2W^\pm)$ & -                  & $5.7\times10^{-3}$ & -      \\
B$(H^\pm\rightarrow JW^\pm)$   & $3.0\times10^{-3}$ & $5.1\times10^{-4}$ & $1.6\times10^{-3}$ \\
B$(H^\pm\rightarrow Z W^\pm)$  & $0.2$              & $0.2$              & $0.3$ \\
\hline\hline
\end{tabular*}
\end{table}
Note that the decay $H^\pm\rightarrow h_2W^\pm$ is not kinematically allowed in B1 and B3. Branching fractions
of $H^\pm\rightarrow h_1W^\pm$ are dominant in the three benchmarks.
%
\begin{figure*}
\includegraphics[width=\textwidth]{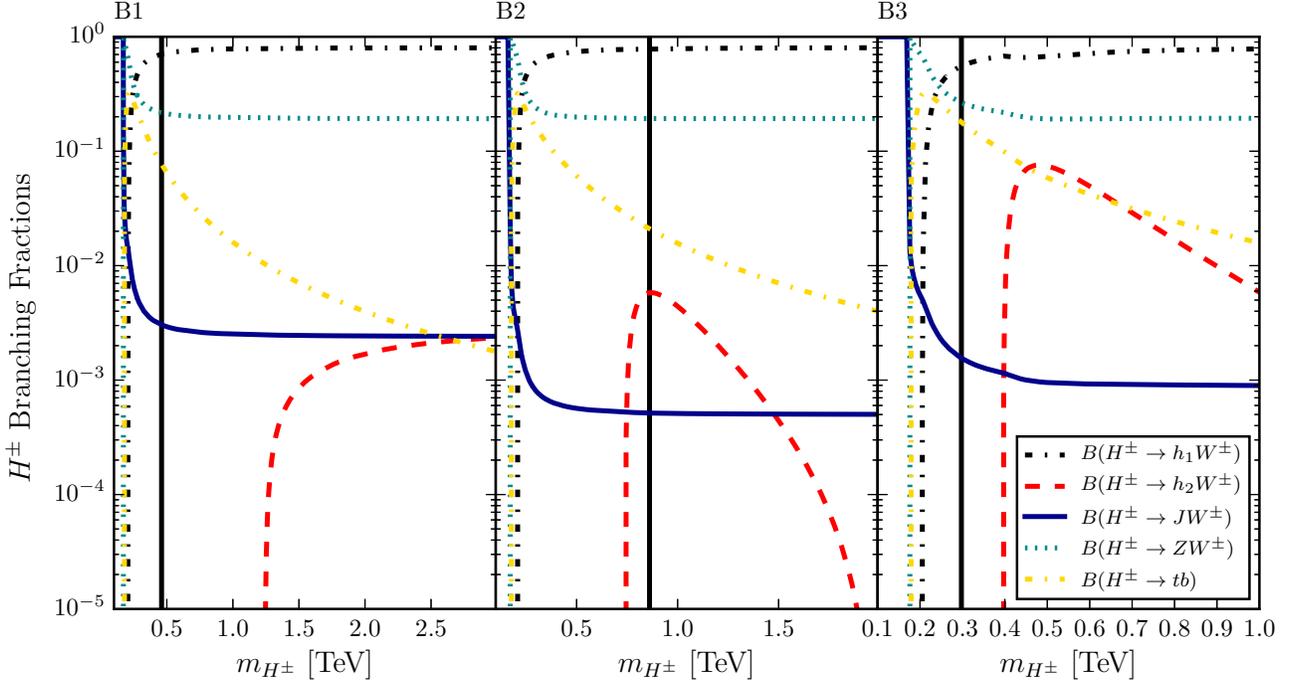}
\caption{Branching fraction for the $H^{+}$ scalar in the three benchmarks as a 
function of $m_{H^{+}}$. The parameter $\kappa$ is varied to 
move $m_{H^{+}}$, as explained in the text. The vertical solid line 
in each frame corresponds to our benchmark point. The plot includes all constrains 
from Section \ref{ParamRestrictions}.
}
\label{BRAVsHP_Curves}
\end{figure*}
%

Fig.~\ref{BRAVsHP_Curves} shows the branching fractions of $H^\pm$ as a function of its mass. 
The curves are obtained starting from each of the 3 benchmarks and vary $\kappa$ 
according to Table \ref{tab1} to change the value of $m_H^\pm$. $\lambda_1$ also varies as in 
Table \ref{tab1} to recover $m_{h_{1}}\approx 125$ GeV.
%
\begin{figure*}
\includegraphics[width=\textwidth]{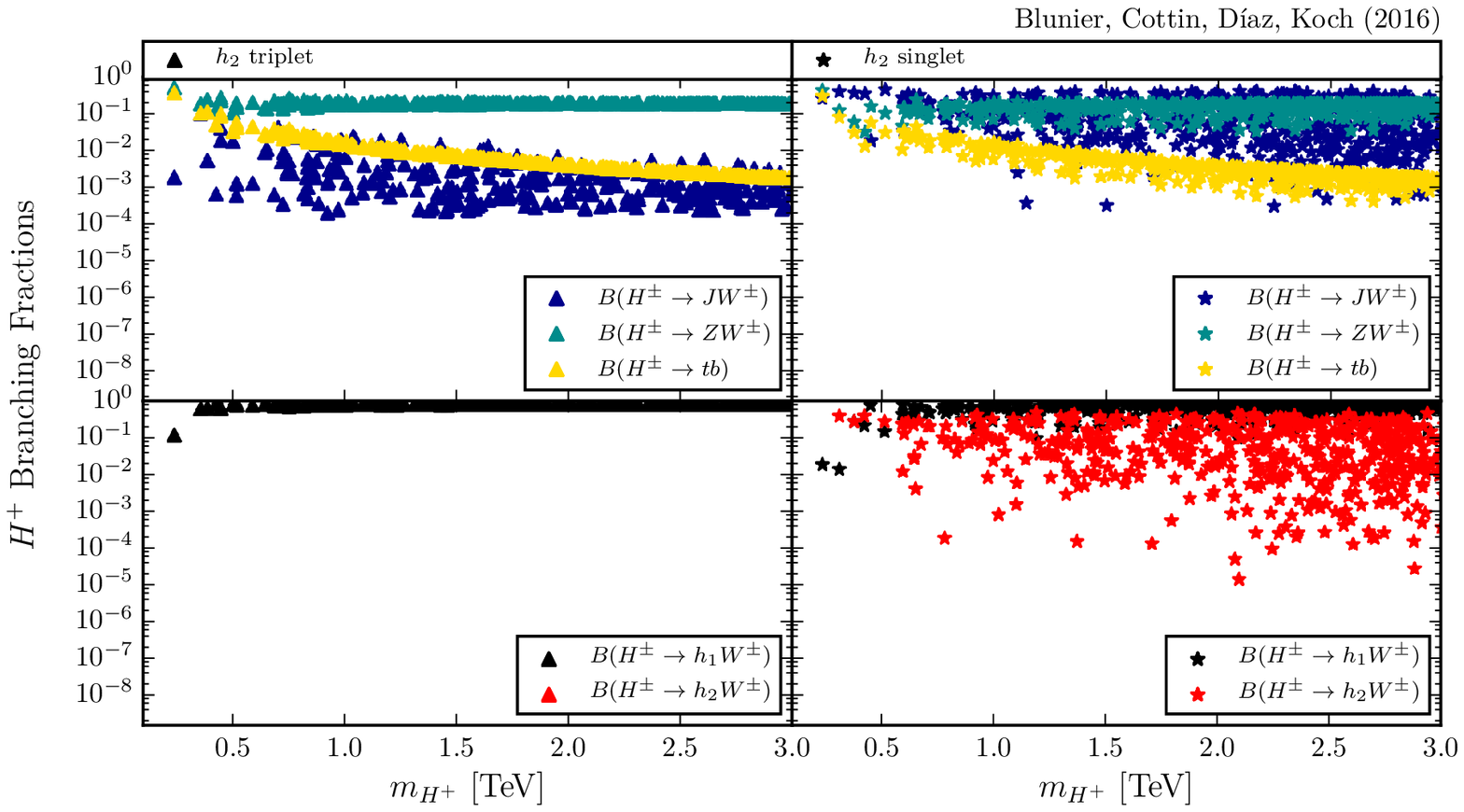}
\caption{Branching fractions for the $H^{+}$ scalar as a function of $m_{H^{+}}$. 
The left column shows points where $h_{2}$ is 
triplet-like (i.e $O^{21}_{\chi}>0.95$). The right column shows 
points where $h_{2}$ is singlet-like (i.e $O^{23}_{\chi}>0.95$). Parameters are 
varied according to Table \ref{tab1}. The scan includes all constrains from 
Section \ref{ParamRestrictions}.
}
\label{BRHPVsmHP_scan}
\end{figure*}
%

Fig.~\ref{BRHPVsmHP_scan} shows the $H^\pm$ branching fractions as a function of its mass
in a general scan. Decays to $h_1W^{\pm}$ dominate, independent of the composition of $h_2$.
Decays to $ZW^{\pm}$ follow in importance. Also important are decays to $h_{2}W^{\pm}$, when $h_{2}$ is 
singlet-like, as when $h_{2}$ is triplet-like, its mass is very close to the mass of $m_{H^{\pm}}$ (as in B1),
so there is no phase space for the decay in this case.
\section{Promising Channels for $h_{2}$, $A$ and $H^{\pm}$}
\label{Channels}

We now briefly comment on the most promising channels for 
discovery of $h_2$, $A$ and $H^{\pm}$ at future $e^{+}e^{-}$ colliders.

A promising channel for the discovery of $h_{2}$, given its large cross-section as 
discussed in Section~\ref{h2Prod}, is $e^{+}e^{-}\rightarrow h_{2}t\bar{t}$. Thinking of B1,
the largest decays fractions for $h_2$ are to $ZZ$ as shown in Table~\ref{tab5}. Considering 
leptonic decays of the $W$ and $Z$, the signal is

\begin{equation}
e^{+}e^{-}\rightarrow ZZ t\bar{t} \rightarrow l^{+}l^{-} l^{+}l^{-} l^{+} \nu_{l} l^{-}\nu_{l} b\bar{b}
\end{equation}

with $l=e,\mu$. The signal contains 2 $b-$jets + 6 leptons + $\ensuremath{p_T^{\mathrm{\, miss}}}$ (missing 
transverse momenta). For B1 at $\sqrt{s}=1$ TeV, the cross-section is estimated as

\begin{align}
\sigma_{2b6l\ensuremath{p_T^{\mathrm{\, miss}}}}&\approx\sigma(e^{+}e^{-}\rightarrow h_2t\bar{t})\times B(h_2\rightarrow ZZ)\nonumber\\
&\times B(Z\rightarrow l^{+}l^{-})^2\times B(W^{\pm}\rightarrow l^{\pm}\nu)^2\nonumber\\
&\approx 3\times 10^{-5} \hspace{0.1cm}\mbox{fb}\label{2b6lsignature}
\end{align}

resulting in less than one event to be discoverable with $\mathcal{L}=1000$ fb$^{-1}$, so 
too little to be observed unfortunately. Possible
SM backgrounds to this signature include $e^{+}e^{-}\rightarrow ZZZ$ and $e^{+}e^{-}\rightarrow ZZt\bar{t}$. Multi-lepton
signatures in the ``23'' HTM were studied in the context of the LHC in Refs.~\cite{Akeroyd:2012nd,delAguila:2008cj}, where
it was shown that after requiring kinematic cuts in the transverse momenta of the leptons, signatures with 6 leptons have no background, even though the signal is also scarce. Therefore, multi-lepton signatures are relevant for higher 
integrated luminosities. We could require similar leptonic kinematic cuts in the case of $e^{+}e^{-}$, in addition 
of requiring 2 $b-$tagged jets and small $\ensuremath{p_T^{\mathrm{\, miss}}}$ due to the two neutrinos. 

For B2 the decay $h_2\rightarrow JJ$ dominates. If one $W$ boson decays hadronically and the
other leptonically, then we will have a 4 $b-$jets + $\ensuremath{p_T^{\mathrm{\, miss}}}$ 
signature, assuming the lepton escapes undetected. This channel was studied in detail in Ref.~\cite{Diaz:1998zg}
for our ``123" model, where it was shown that with appropriate cuts in $\ensuremath{p_T^{\mathrm{\, miss}}}$, number
of jets and invariant mass distributions the background is removed while keeping high signal efficiency.

In the case of the $CP-$odd Higgs $A$, there are two relevant processes. $e^{+}e^{-}\rightarrow AZZ$ has 
the highest cross-section for B1 and B2. In the case where $A\rightarrow t\bar{t}$ we have the same signature as
before for $h_{2}$. The decay $A\rightarrow h_{1}Z$ also dominates in our benchmarks. The dominant decay $h_{1}\rightarrow b\bar{b}$ 
follows, leading to topologies with leptons and $b-$jets (with no missing transverse momenta), depending on the decay of
the $Z$. The cross-section for, 

\begin{equation}
e^+e^-\rightarrow A ZZ\rightarrow h_1 ZZZ \rightarrow b\bar{b}l^{+}l^{-}l^{+}l^{-}l^{+}l^{-}  
\end{equation}

leads to a 2$b-$jet+6 leptons signature. The cross-section for B1 at $\sqrt{s}=1$ TeV is estimated as,

\begin{align}
\sigma_{2b6l}&\approx \sigma(e^{+}e^{-}\rightarrow A ZZ) \times B(A\rightarrow h_{1}Z)\times B(h_{1}\rightarrow b\bar{b}) \nonumber\\
&\times B(Z\rightarrow l^{+}l^{-})^{3}\nonumber\\
&\approx 1.0\times10^{-4}\hspace{0.1cm}\mbox{fb}
\end{align}

resulting in less than one event with $\mathcal{L}=1000$ fb$^{-1}$. Possible backgrounds
are very similar and include the ones in equation~\ref{2b6lsignature}, so similar cuts can be applied to suppress them. 

The associated production $e^{+}e^{-}\rightarrow AJJ$ dominates in B3 with $A\rightarrow b\bar{b}$, leading
to the topology of 2 $b-$jets + $\ensuremath{p_T^{\mathrm{\, miss}}}$. This signal 
was studied for the ``23'' HTM in~\cite{deCampos:1996bg}, with largest background coming from 
$e^{+}e^{-}\rightarrow W^{+}W^{-}$ and $e^{+}e^{-}\rightarrow ZZ$. The authors concluded 
that the most efficient way to improve the signal-to-background ratio is to require $b-$tagged jets and 
large $\ensuremath{p_T^{\mathrm{\, miss}}}$, in addition to charged multiplicity and
an invariant mass cut close to the mass of the visibly decaying particle.

Production for the singly charged Higgs dominates in  
$e^+e^-\rightarrow H^{+}H^{-}\rightarrow H^+h_1W^-$ for most 
of our benchmarks (see Figure~\ref{Hp_prod}). This is followed by
the decay of $H^{+}\rightarrow h_{1} W^{+}$, which has the highest 
branching fraction (see Table~\ref{tab7}). An optimal discovery channel would be
when $h_{1}\rightarrow b\bar{b}$ and when one $W$ boson decays hadronically and the other 
leptonically,

\begin{equation}
e^+e^-\rightarrow H^+h_1W^-\rightarrow h_{1} W^{+}h_1W^-\rightarrow b\bar{b} l^{\pm}\nu_{l}b\bar{b}q\bar{q}
\end{equation}

resulting in an event topology of $4b-$jets + 2 jets + 1 lepton + $\ensuremath{p_T^{\mathrm{\, miss}}}$, where 
the lepton $l=e,\mu$. This distinctive signature was 
studied for a charged Higgs in the context of Two-Higgs doublet models~\cite{Moretti:2003cd,Komamiya:1988rs}. The 
mass of the singly charged Higgs can be reconstructed and the events 
can be selected with $b-$tagging techniques, in addition to requiring one isolated 
lepton. Also, two jets must have the $W$ mass. 

We can estimate the visible cross-section for this final state. For $\sqrt{s}=1$ TeV 
in B1 we have,

\begin{align}
\sigma_{4b\ensuremath{p_T^{\mathrm{\, miss}}}ljj} &\approx 
\sigma(e^{+}e^{-}\rightarrow H^{+}h_1W^{-})\nonumber\\
&\times B(H^{+}\rightarrow h_1 W^{+})\times B(h_{1}\rightarrow b\bar{b})^2\nonumber\\
&\times B(W^{\pm}\rightarrow l^{\pm}\nu_{l})\times B(W^{\pm}\rightarrow q\bar{q})\nonumber\\
&\approx 0.04 \hspace{0.1cm}\mbox{fb}
\end{align}

and since the ILC has a yearly integrated luminosity 
of $1000$ fb$^{-1}$, this results in about 40 potentially discoverable events. A relevant 
SM background for this signature is
the process $e^{+}e^{-}\rightarrow t\bar{t}b\bar{b}$. Our estimation 
yields a visible cross-section 
of $\sigma_{\mbox{\tiny{SM-}}4b\ensuremath{p_T^{\mathrm{\, miss}}}ljj}\approx 0.4$ fb, which
is quite significant. The signal-to-background ratio can be enhanced by applying the selection 
cuts above mentioned. It was also shown in Ref.~\cite{Moretti:2003cd} that one can suppress this big irreducible background 
to a negligible level by using a technique that allows the reconstruct of the neutrino four-momentum.

Of course a more detailed simulation study should be done in order to suppress backgrounds 
further and improve signal efficiency for the channels mentioned. A fully fledge 
study in this direction, considering also detector efficiencies, goes beyond 
the scope of this paper and we leave it for a future work.

\section{Conclusions}
\label{close}

We have studied the Higgs phenomenology of a model with a scalar triplet, a scalar singlet
and a scalar doublet under $SU(2)$. In this ``123'' variant of the Higgs triplet model
the singlet acquires a vacuum expectation value, which spontaneously breaks lepton
number. The vacuum expectation value generated for the triplet provides a mass 
term for neutrinos. This feature makes it a well motivated model to look for at particle colliders.

The lightest $CP-$even Higgs, $h_{1}$, has been identified with the SM-like Higgs 
boson discovered at the LHC, which constrains the parameters in the scalar
potential of the model. We studied the production cross-sections and decay ratios of the second 
heaviest $CP-$even Higgs $h_{2}$, the $CP-$odd Higgs $A$ and the singly charged Higgs $H^{\pm}$. We found 
that production cross-sections at hadron colliders can be very low for these states, so we perform a numerical 
analysis assessing the discovery potential at future lepton colliders. 

We find characteristic features in cases where $h_{2}$ is singlet-like, triplet-like or a mixture. The 
main 2-body production mode for $h_{2}$ is associated production with a $CP-$odd state $A$. We note 
that cross-sections for $A$ and $H^{\pm}$ are enhanced when a second heavy particle is also 
produced on-shell. Invisible decays of $h_{2}$ to Majorons can be very important. Decays 
of the singly charged Higgs $H^{\pm}\rightarrow h_{1}W^{\pm}$ dominate. These features lead 
to promising channels for discovery of $h_2$ and $A$, in particular in the 
$4b-$jets+$\ensuremath{p_T^{\mathrm{\, miss}}}$ and $2b-$jets+$\ensuremath{p_T^{\mathrm{\, miss}}}$ final states, as shown in Ref.~\cite{Diaz:1998zg} and Ref.~\cite{deCampos:1996bg}, respectively, as we estimate the most 
promising signal with leptons in the final state are too small to be observed. The 
$4b-$jets + 2 jets + 1 lepton + $\ensuremath{p_T^{\mathrm{\, miss}}}$ final state is optimal 
for the discovery of the singly charged Higgs. These signals provides a test of the ``123" HTM at future $e^{+}e^{-}$ colliders.

\begin{acknowledgments}
SB was funded by the PUC Vice Rectory of Research Scholarship. The work of MAD was 
supported by Fondecyt 1141190. The work of BK was supported by 
Fondecyt 1161150. GC was funded by the postgraduate 
Conicyt-Chile/Cambridge Trusts Scholarship 84130011 and also acknowledges partial 
support from STFC grant ST/L000385/1. 
\end{acknowledgments}
\appendix
\section{Convention for Diagonalization.}
\label{Diagonalization}

The diagonalization in the charged scalar sector is,
\begin{equation}
\left[ \begin{array}{c} h^+_1 \\ h^+_2 \end{array} \right] \equiv
\left[ \begin{array}{c} G^+ \\ H^+ \end{array} \right] = O_+
\left[ \begin{array}{c} \phi^{-*} \\ \Delta^{+} \end{array} \right] \equiv
\left( \begin{array}{cc} - c_\beta & s_\beta \\ s_\beta & c_\beta \end{array} \right)
\left[ \begin{array}{c} \phi^{-*} \\ \Delta^{+} \end{array} \right]
\end{equation}
and the diagonalization in the neutral scalar sector proceeds as,
\begin{equation}
\left[ \begin{array}{c} h_1 \\ h_2 \\ h_3 \end{array} \right] =
O_\chi
\left[ \begin{array}{c} \chi_\sigma \\ \chi_\phi \\ \chi_\Delta \end{array} \right] \,,\qquad
\left[ \begin{array}{c} a_1 \\ a_2 \\ a_3 \end{array} \right] \equiv
\left[ \begin{array}{c} G^0 \\ J \\ A \end{array} \right] =
O_\varphi
\left[ \begin{array}{c} \varphi_\sigma \\ \varphi_\phi \\ \varphi_\Delta \end{array} \right],
\end{equation}
where $O_\chi$ and $O_\varphi$ are $3\times3$ matrices.

The mass matrix in eq.~(\ref{CPoddScalars}) is diagonalized by the matrix,
\begin{equation}
O_\varphi = \left[ \begin{array}{ccc} 
0 & \frac{1}{N_G} & - \frac{2}{N_G} \frac{v_\Delta}{v_\phi} \\
\frac{N^2_G}{N_J}  & - \frac{2}{N_J} \frac{v_\Delta^2}{v_\phi v_\sigma} & - 
\frac{1}{N_J} \frac{v_\Delta}{v_\sigma} \\
\frac{1}{N_A} \frac{v_\Delta}{v_\sigma} & \frac{2}{N_A} \frac{v_\Delta}{v_\phi} & \frac{1}{N_A}
\end{array}
\right],
\end{equation}
where
\begin{eqnarray}
N_G &=& \sqrt{ 1 + 4 \frac{v_\Delta^2}{v_\phi^2} },
\nonumber\\
N_J &=& \sqrt{ N^4_G + 4 \frac{v_\Delta^4}{v_\phi^2 v_\sigma^2} + \frac{v_\Delta^2}{v_\sigma^2}},
\nonumber\\
N_A &=& \sqrt{1+4\frac{v_\Delta^2}{v_\phi^2}+\frac{v_\Delta^2}{v_\sigma^2}}.
\end{eqnarray}
The mass eigenstate fields are,
\begin{eqnarray}
G^0 &=& \frac{1}{N_G} \varphi_\phi - \frac{2}{N_G} \frac{v_\Delta}{v_\phi} \varphi_\Delta,
\nonumber\\
J &=& \frac{N^2_G}{N_J} \varphi_\sigma - \frac{2}{N_J} \frac{v_\Delta^2}{v_\phi v_\sigma} \varphi_\phi
- \frac{1}{N_J} \frac{v_\Delta}{v_\sigma} \varphi_\Delta,
\nonumber\\
A &=& \frac{1}{N_A} \frac{v_\Delta}{v_\sigma} \varphi_\sigma 
+ \frac{2}{N_A} \frac{v_\Delta}{v_\phi} \varphi_\phi + \frac{1}{N_A} \varphi_\Delta.
\label{NeuFields123Model}
\end{eqnarray}
From here we conclude that the Majoron has the tendency to be mainly singlet and that the neutral Goldstone boson 
has no singlet component (the singlet does not couple to the $Z$ boson).

\section{Feynman Rules.}
\label{FeynmanRules}

\subsection{One scalar and two fermions}

%
\begin{center}
\vspace{-50pt} \hfill \\
\begin{axopicture}(100,90)(10,23) 
\DashLine(0,30)(40,30){5}
\ArrowLine(70,60)(40,30)
\ArrowLine(40,30)(70,0)
\Text(5,40)[]{$h_i$}
\Text(80,60)[]{$\bar{f}$}
\Text(80,2)[]{$f$}
\end{axopicture}
\vspace{30pt}
$=-iO_\chi^{i2}\frac{h_f}{\sqrt{2}}$ \hfill
\end{center}
\begin{center}
\vspace{-50pt} \hfill \\
\begin{axopicture}(100,90)(10,23) 
\DashLine(0,30)(40,30){5}
\ArrowLine(70,60)(40,30)
\ArrowLine(40,30)(70,0)
\Text(5,40)[]{$a_i$}
\Text(80,60)[]{$\bar{f}$}
\Text(80,2)[]{$f$}
\end{axopicture}
\vspace{30pt}
$=O_\varphi^{i2}\frac{h_f}{\sqrt{2}}\gamma_5$ \hfill
\end{center}
%

\subsection{One scalar and two gauge bosons}

%
\begin{center}
\vspace{-50pt} \hfill \\
\begin{axopicture}(100,90)(10,23) 
\DashLine(0,30)(40,30){5}
\Photon(70,60)(40,30){3}{5}
\Photon(70,0)(40,30){-3}{5}
\Text(5,40)[]{$h_i$}
\Text(80,60)[]{$Z^\mu$}
\Text(80,2)[]{$Z^\nu$}
\end{axopicture}
\vspace{30pt}
$=i \frac{1}{2}(g^2+g'^2) (O_\chi^{i2}v_\phi+4O_\chi^{i3}v_\Delta) g^{\mu\nu}$ \hfill
\end{center}
\begin{center}
\vspace{-50pt} \hfill \\
\begin{axopicture}(100,90)(10,23) 
\DashLine(0,30)(40,30){5}
\Photon(70,60)(40,30){3}{5}
\Photon(70,0)(40,30){-3}{5}
\Text(5,40)[]{$h_i$}
\Text(85,60)[]{$W^+_\mu$}
\Text(85,2)[]{$W^-_\nu$}
\end{axopicture}
\vspace{30pt}
$=i\frac{g^2}{2}(O_\chi^{i2}v_\phi+2O_\chi^{i3}v_\Delta) g_{\mu\nu}$ \hfill
\end{center}
%

\subsection{Two scalars and one gauge boson}

%
\begin{center}
\vspace{-50pt} \hfill \\
\begin{axopicture}(100,90)(10,23) 
\Photon(0,30)(40,30){2}{5}
\DashArrowLine(70,60)(40,30){5}
\DashArrowLine(70,0)(40,30){5}
\LongArrow(50,50)(55,55)
\LongArrow(55,5)(50,10)
\Text(10,40)[]{$Z_\mu$}
\Text(80,60)[]{$h_i$}
\Text(80,2)[]{$a_j$}
\Text(45,47)[]{$p'$}
\Text(45,13)[]{$p$}
\end{axopicture}
\vspace{30pt} 
$=\frac{g}{2c_W}(O_\chi^{i2}O_\varphi^{j2}-2O_\chi^{i3}O_\varphi^{j3})(p+p')_\mu$ \hfill
\end{center}
\begin{center}
\vspace{-40pt} \hfill \\
\begin{axopicture}(70,70)(40,27) 
\Photon(0,30)(40,30){2}{5}
\DashArrowLine(70,60)(40,30){5}
\DashArrowLine(70,0)(40,30){5}
\LongArrow(50,50)(55,55)
\LongArrow(55,5)(50,10)
\Text(5,40)[]{$Z_\mu$}
\Text(82,60)[]{$h_i^-$}
\Text(82,2)[]{$h_j^+$}
\Text(45,47)[]{$p'$}
\Text(45,13)[]{$p$}
\end{axopicture}
\vspace{30pt} 
$=-\frac{ig}{2c_W}\Big[O_+^{i1}O_+^{j1}(c_W^2-s_W^2)-2O_+^{i2}O_+^{j2}s_W^2\Big](p+p')_\mu$ \hfill
\end{center}
\vspace{10pt}
\begin{center}
\vspace{-40pt} \hfill \\
\begin{axopicture}(70,70)(40,27) 
\Photon(0,30)(40,30){2}{5}
\DashArrowLine(70,60)(40,30){5}
\DashArrowLine(70,0)(40,30){5}
\LongArrow(50,50)(55,55)
\LongArrow(55,5)(50,10)
\Text(5,40)[]{$Z_\mu$}
\Text(90,60)[]{$\Delta^{++*}$}
\Text(87,2)[]{$\Delta^{++}$}
\Text(45,47)[]{$p'$}
\Text(45,13)[]{$p$}
\end{axopicture}
\vspace{30pt} 
$=-\frac{ig}{c_W}(c_W^2-s_W^2)(p+p')_\mu$ \hfill
\end{center}
\vspace{10pt}
\begin{center}
\vspace{-50pt} \hfill \\
\begin{axopicture}(100,90)(10,23) 
\Photon(0,30)(40,30){2}{5}
\DashArrowLine(70,60)(40,30){5}
\DashArrowLine(70,0)(40,30){5}
\LongArrow(50,50)(55,55)
\LongArrow(55,5)(50,10)
\Text(15,40)[]{$A_\mu$}
\Text(84,60)[]{$h^-_i$}
\Text(84,2)[]{$h^+_j$}
\Text(45,47)[]{$p'$}
\Text(45,13)[]{$p$}
\end{axopicture}
\vspace{30pt} 
$=-ie(p+p')_\mu \delta_{ij}$ \hfill
\end{center}
\begin{center}
\vspace{-50pt} \hfill \\
\begin{axopicture}(100,90)(10,23) 
\Photon(0,30)(40,30){2}{5}
\DashArrowLine(70,60)(40,30){5}
\DashArrowLine(70,0)(40,30){5}
\LongArrow(50,50)(55,55)
\LongArrow(55,5)(50,10)
\Text(15,40)[]{$A_\mu$}
\Text(88,60)[]{$\Delta^{++*}$}
\Text(84,2)[]{$\Delta^{++}$}
\Text(45,47)[]{$p'$}
\Text(45,13)[]{$p$}
\end{axopicture}
\vspace{30pt} 
$=-2ie(p+p')_\mu$ \hfill
\end{center}
\begin{center}
\vspace{-50pt} \hfill \\
\begin{axopicture}(100,90)(10,23) 
\Photon(0,30)(40,30){2}{5}
\DashArrowLine(70,60)(40,30){5}
\DashArrowLine(70,0)(40,30){5}
\LongArrow(50,50)(55,55)
\LongArrow(55,5)(50,10)
\Text(15,40)[]{$W_\mu^+$}
\Text(84,60)[]{$h^-_i$}
\Text(84,2)[]{$h_j$}
\Text(45,47)[]{$p'$}
\Text(45,13)[]{$p$}
\end{axopicture}
\vspace{30pt} 
$=i\frac{g}{2}\big(O_+^{i1}O_\chi^{j2}-\sqrt{2}O_+^{i2}O_\chi^{j3}\big)(p+p')_\mu$ \hfill
\end{center}
\vspace{10pt} 
\begin{center}
\vspace{-50pt} \hfill \\
\begin{axopicture}(100,90)(10,23) 
\Photon(0,30)(40,30){2}{5}
\DashArrowLine(70,60)(40,30){5}
\DashArrowLine(70,0)(40,30){5}
\LongArrow(50,50)(55,55)
\LongArrow(55,5)(50,10)
\Text(15,40)[]{$W_\mu^+$}
\Text(84,60)[]{$h^-_i$}
\Text(84,2)[]{$a_j$}
\Text(45,47)[]{$p'$}
\Text(45,13)[]{$p$}
\end{axopicture}
\vspace{30pt} 
$=i\frac{g}{2}\big(O_+^{i1}O_\chi^{j2}+\sqrt{2}O_+^{i2}O_\chi^{j3}\big)(p+p')_\mu$ \hfill
\end{center}
\vspace{10pt} 

\subsection{Three Scalars}
For the case with one $CP-$even and two $CP-$odd Higgs bosons, the relevant term in the Lagrangian is
\begin{eqnarray}
{\cal L}_{h_i a_j a_k} &=& M_{h_i a_j a_k} h_i a_j a_k,
\end{eqnarray}
where we sum over $i,j,k$. The coupling $M_{h_i a_j a_k}$ (with 
units of mass), after symmetrization in $j$ and $k$ is given by the expression
\begin{widetext}
\begin{eqnarray}
M_{h_i a_j a_k} &=&
- \lambda_1 v_\phi O_\chi^{i2} O_\varphi^{j2} O_\varphi^{k2}
- (\lambda_2+\lambda_4) v_\Delta O_\chi^{i3} O_\varphi^{j3} O_\varphi^{k3}
- \frac{1}{2} ( \lambda_3 + \lambda_5 ) v_\phi O_\chi^{i2} O_\varphi^{j3} O_\varphi^{k3}
\nonumber\\ &&
- \, \frac{1}{2} \big[ ( \lambda_3 + \lambda_5 ) v_\Delta + \kappa v_\sigma \big]
O_\chi^{i3} O_\varphi^{j2} O_\varphi^{k2}
- \beta_1 v_\sigma O_\chi^{i1} O_\varphi^{j1} O_\varphi^{k1}
- \frac{1}{2} \beta_2 v_\phi O_\chi^{i2} O_\varphi^{j1} O_\varphi^{k1}
\nonumber\\ &&
- \, \frac{1}{2} ( \beta_2 v_\sigma + \kappa v_\Delta ) O_\chi^{i1} O_\varphi^{j2} O_\varphi^{k2}
- \frac{1}{2} \beta_3 v_\Delta O_\chi^{i3} O_\varphi^{j1} O_\varphi^{k1}
- \frac{1}{2} \beta_3 v_\sigma O_\chi^{i1} O_\varphi^{j3} O_\varphi^{k3}
\nonumber\\ &&
- \, \frac{1}{2} \kappa v_\phi O_\chi^{i2} (O_\varphi^{j1} O_\varphi^{k3}+O_\varphi^{k1} O_\varphi^{j3})
- \frac{1}{2} \kappa v_\phi O_\chi^{i3} (O_\varphi^{j1} O_\varphi^{k2}+O_\varphi^{k1} O_\varphi^{j2})
- \frac{1}{2} \kappa v_\phi O_\chi^{i1} (O_\varphi^{j2} O_\varphi^{k3}+O_\varphi^{k2} O_\varphi^{j3})
\nonumber\\ &&
- \, \frac{1}{2} \kappa v_\Delta O_\chi^{i2} (O_\varphi^{j1} O_\varphi^{k2}+O_\varphi^{k1} O_\varphi^{j2})
- \frac{1}{2} \kappa v_\sigma O_\chi^{i2} (O_\varphi^{j2} O_\varphi^{k3}+O_\varphi^{k2} O_\varphi^{j3}).
\end{eqnarray}
\end{widetext}
This leads to the following Feynman rule,
\begin{center}
\vspace{-55pt} \hfill \\
\begin{axopicture}(100,90)(10,23) 
\DashLine(0,30)(40,30){5}
\DashLine(40,30)(70,60){5}
\DashLine(40,30)(70,0){5}
\Text(5,40)[]{$h_i$}
\Text(80,60)[]{$a_j$}
\Text(80,2)[]{$a_k$}
\end{axopicture}
\vspace{30pt}
\hspace{-0.3cm}$=iM_{h_i a_j a_k}$ (twice larger if $j=k$). \hfill
\end{center}
For one $CP-$even and two charged Higgs bosons, the relevant term in the Lagrangian is,
\begin{eqnarray}
{\cal L}_{h_i h^+_j h^-_k} &=& M_{h_i h^+_j h^-_k} h_i h^+_j h^-_k
\end{eqnarray}
where we sum over $i,j,k$. The coupling $M_{h_i h^+_j h^-_k}$ (with 
units of mass) is given by the expression
\begin{widetext}
\begin{eqnarray}
M_{h_i h_j^+ h_k^-} &=&
- 2 \lambda_1 v_\phi O_\chi^{i2} O_+^{j1} O_+^{k1}
- 2 (\lambda_2+\lambda_4) v_\Delta O_\chi^{i3} O_+^{j2} O_+^{k2}
- ( \lambda_3 + \frac{1}{2} \lambda_5 ) v_\phi O_\chi^{i2} O_+^{j2} O_+^{k2}
\nonumber\\ &&
- \lambda_3 v_\Delta O_\chi^{i3} O_+^{j1} O_+^{k1}
- \frac{1}{2\sqrt{2}} \lambda_5 v_\phi O_\chi^{i3} O_+^{j2} O_+^{k1}
- \frac{1}{2\sqrt{2}} \lambda_5 v_\phi O_\chi^{i3} O_+^{j1} O_+^{k2}
\nonumber\\ &&
- \frac{1}{\sqrt{2}} ( {\textstyle{\frac{1}{2}}} \lambda_5 v_\Delta - \kappa v_\sigma)
O_\chi^{i2} O_+^{j2} O_+^{k1}
- \frac{1}{\sqrt{2}} ( {\textstyle{\frac{1}{2}}} \lambda_5 v_\Delta - \kappa v_\sigma)
O_\chi^{i2} O_+^{j1} O_+^{k2}
- \beta_2 v_\sigma O_\chi^{i1} O_+^{j1} O_+^{k1}
\nonumber\\ &&
- \beta_3 v_\sigma O_\chi^{i1} O_+^{j2} O_+^{k2}
+ \frac{1}{\sqrt{2}} \kappa v_\phi O_\chi^{i1} O_+^{j2} O_+^{k1}
+ \frac{1}{\sqrt{2}} \kappa v_\phi O_\chi^{i1} O_+^{j1} O_+^{k2}
\end{eqnarray}
\end{widetext}
and the Feynman rule is,
\begin{center}
\vspace{-50pt} \hfill \\
\begin{axopicture}(100,90)(10,23) 
\DashLine(0,30)(40,30){5}
\DashArrowLine(40,30)(70,60){5}
\DashArrowLine(70,0)(40,30){5}
\Text(5,40)[]{$h_i$}
\Text(80,60)[]{$h^+_j$}
\Text(80,2)[]{$h^-_k$}
\end{axopicture}
\vspace{30pt}
$=iM_{h_i h^+_j h^-_k}$.\hfill
\end{center}
For one $CP-$even and two doubly charged Higgs bosons, the relevant term in the Lagrangian is
\begin{equation}
{\cal L}_{h_i \Delta^{++}\Delta^{--}} = M_{h_i \Delta^{++} \Delta^{--}} h_i \Delta^{++*} \Delta^{++},
\end{equation}
with
\begin{equation}
M_{h_i \Delta^{++} \Delta^{--}} = - 2 \lambda_2 v_\Delta O_\chi^{i3} - \lambda_3 v_\phi O_\chi^{i2}
- \beta_3 v_\sigma O_\chi^{i1},
\end{equation}
leading to the following Feynman rule
\begin{center}
\vspace{-50pt} \hfill \\
\begin{axopicture}(100,90)(10,23) 
\DashLine(0,30)(40,30){5}
\DashArrowLine(40,30)(70,60){5}
\DashArrowLine(70,0)(40,30){5}
\Text(5,40)[]{$h_i$}
\Text(90,60)[]{$\Delta^{++}$}
\Text(90,2)[]{$\Delta^{--}$}
\end{axopicture}
\vspace{30pt}
$=iM_{h_i \Delta^{++} \Delta^{--}}$. \hfill
\end{center}
For three $CP-$even Higgs bosons, the relevant term in the Lagrangian is 
\begin{eqnarray}
{\cal L}_{h_i h_j h_k} &=& M_{h_i h_j h_k} h_i h_j h_k ,
\end{eqnarray}
where we sum over $i,j,k$. The coupling $M_{h_i h_j h_k}$ (with units of mass), after symmetrization in $j$ and $k$, is given by
\begin{widetext}
\begin{eqnarray}
M_{h_i h_j h_k} &=& 
- 6 \lambda_1 v_\phi O_\chi^{i2} O_\chi^{j2} O_\chi^{k2}
- 6 (\lambda_2+\lambda_4) v_\Delta O_\chi^{i3} O_\chi^{j3} O_\chi^{k3}
\nonumber\\ &&
- \, ( \lambda_3 + \lambda_5 ) v_\phi \Big[ O_\chi^{i2} O_\chi^{j3} O_\chi^{k3}+
O_\chi^{k2} O_\chi^{i3} O_\chi^{j3} + O_\chi^{j2} O_\chi^{k3} O_\chi^{i3} \Big]
\nonumber\\ &&
- \, \big[ ( \lambda_3 + \lambda_5 ) v_\Delta - \kappa v_\sigma \big]
\Big[ O_\chi^{i2} O_\chi^{j2} O_\chi^{k3} + O_\chi^{k2} O_\chi^{i2} O_\chi^{j3} +
O_\chi^{j2} O_\chi^{k2} O_\chi^{i3} \Big]
- 6 \beta_1 v_\sigma O_\chi^{i1} O_\chi^{j1} O_\chi^{k1} 
\nonumber\\ &&
- \, \beta_2 v_\phi \Big[ O_\chi^{i1} O_\chi^{j1} O_\chi^{k2} + 
O_\chi^{k1} O_\chi^{i1} O_\chi^{j2} + O_\chi^{j1} O_\chi^{k1} O_\chi^{i2} \Big]
\nonumber\\ &&
- \, ( \beta_2 v_\sigma - \kappa v_\Delta) 
\Big[ O_\chi^{i1} O_\chi^{j2} O_\chi^{k2} + O_\chi^{k1} O_\chi^{i2} O_\chi^{j2} +
O_\chi^{j1} O_\chi^{k2} O_\chi^{i2} \Big]
\nonumber\\ &&
- \, \beta_3 v_\Delta \Big[ O_\chi^{i1} O_\chi^{j1} O_\chi^{k3} +
O_\chi^{k1} O_\chi^{i1} O_\chi^{j3} + O_\chi^{j1} O_\chi^{k1} O_\chi^{i3}\Big]
\nonumber\\ &&
- \, \beta_3 v_\sigma \Big[ O_\chi^{i1} O_\chi^{j3} O_\chi^{k3} +
O_\chi^{k1} O_\chi^{i3} O_\chi^{j3} + O_\chi^{j1} O_\chi^{k3} O_\chi^{i3} \Big]
\nonumber\\ && \!\!\!\!\!\!\!\!\!\!\!\!\!\!\!\!
+ \, \kappa v_\phi \Big[ O_\chi^{i1} O_\chi^{j2} O_\chi^{k3} + O_\chi^{i1} O_\chi^{k2} O_\chi^{j3} +
O_\chi^{j1} O_\chi^{i2} O_\chi^{k3} + O_\chi^{k1} O_\chi^{i2} O_\chi^{j3} +
O_\chi^{j1} O_\chi^{k2} O_\chi^{i3} + O_\chi^{k1} O_\chi^{j2} O_\chi^{i3} \Big]
\end{eqnarray}
\end{widetext}
The corresponding Feynman rule is given by
\begin{center}
\vspace{-50pt} 
\hfill \\
\begin{axopicture}(100,90)(10,23) 
\DashLine(0,30)(40,30){5}
\DashArrowLine(40,30)(70,60){5}
\DashArrowLine(70,0)(40,30){5}
\Text(5,40)[]{$h_2$}
\Text(90,60)[]{$h_1$}
\Text(90,2)[]{$h_1$}
\end{axopicture}
\vspace{30pt}
$=iM_{h_2h_1h_1}$. \hfill
\end{center}

\bibliographystyle{apsrev4-1}
\bibliography{HiggsTriplet}

\end{document}